\documentstyle[10pt,aaspp4]{article}

\newcounter{romnum}

\def\deg{\circ}

\begin{document}

\title{Estimation of relativistic accretion disk parameters from
iron line emission}

\author{V.~I.~Pariev\altaffilmark{1,2}}
\affil{Theoretical Astrophysics Group, T-6, MS B288, 
Los Alamos National Laboratory, Los Alamos, NM 87545}
\author{B.~C.~Bromley}
\affil{Department of Physics, The University of Utah, 201 James Fletcher
Bldg, Salt Lake City, Utah 84112}
\author{W.~A.~Miller}
\affil{Theoretical Astrophysics Group, T-6, MS B288, 
Los Alamos National Laboratory, Los Alamos, NM 87545}

\altaffiltext{1}{Steward Observatory, University of Arizona, 933 North Cherry
Avenue, Tucson, AZ 85721}
\altaffiltext{2}{P.~N.~Lebedev Physical Institute,
Leninsky Prospect 53, Moscow 117924, Russia}

\begin{abstract}

The observed iron K$\alpha$ fluorescence lines in
Seyfert~\uppercase{i} galaxies provide strong evidence for an
accretion disk near a supermassive black hole as a source of the
emission.  Here we present an analysis of the geometrical and
kinematic properties of the disk based on the extreme frequency shifts
of a line profile as determined by measurable flux in both the red and
blue wings.  The edges of the line are insensitive to the distribution
of the X-ray flux over the disk, and hence provide a robust
alternative to profile fitting of disk parameters.  Our approach
yields new, strong bounds on the inclination angle of the disk and the
location of the emitting region.  We apply our method to interpret
observational data from MCG-6-30-15 and find that the commonly assumed
inclination $30^\deg$ for the accretion disk in MCG-6-30-15 is
inconsistent with the position of the blue edge of the line at a
$3\sigma$ level.  A thick turbulent disk model or the presence of
highly ionized iron may reconcile the bounds on inclination from the
line edges with the full line profile fits based on simple,
geometrically thin disk models.  
The bounds on the innermost radius of disk emission indicate that the black 
hole in MCG-6-30-15 is rotating faster than 30~\% of theoretical 
maximum.
When applied to data from NGC~4151,
our method gives bounds on
the inclination angle of the X-ray emitting inner disk of
$50\pm 10^\deg$, consistent with the presence of an ionization cone
grazing the disk as proposed by Pedlar et al.~(1993).  The frequency extrema
analysis also provides limits to the innermost disk radius in another
Seyfert~1 galaxy, NGC~3516, and is suggestive of a thick disk model.

\end{abstract}


\section{Introduction}

The {\it Advanced Satellite for Cosmology and Astrophysics} (ASCA) has
provided data from over a dozen Seyfert~\uppercase{i} galaxies to
reveal the presence of iron emission lines which are broadened by a
considerable fraction of the speed of light --- greater than 0.2~$c$
in some cases (Mushotzky et al. 1995; Tanaka et al. 1995; Nandra et
al. 1997a). The observed line profiles are the most direct evidence
for the presence of supermassive ($\sim 10^{8}$~M$_\odot$) black holes
in the centers of these galaxies: the spectra have distinctive skewed,
double-peaked profiles which reflect the Doppler and gravitational
shifts associated with emitting material in a strongly curved
spacetime (Chen \& Halpern 1989; Fabian et al. 1989; Laor 1991; Kojima
1991).  The data strongly support a model wherein the emission lines
are produced by iron K$\alpha$ fluorescence at 6.4~keV when optically
thick, ``cold'' regions of an accretion disk (such that the ionization
state of iron is less than Fe~XVII) are externally illuminated by hard
X-rays~(George~\& Fabian 1991; Matt, Perola \& Piro 1991).  In one
bright, well-studied source, MCG-6-30-15, the high signal-to-noise
ratio has enabled parameters of a simple geometrically thin,
relativistic model to be estimated (Tanaka et al. 1995; Dabrowski et
al. 1997; Reynolds, Begelman 1997; Bromley, Miller \& Pariev 1998).

The model parameters which may be gleaned from line profiles include
disk radii, emissivity of the disk, observed inclination angle of the
disk~$i$ ($i=0^\deg$ is face-on and $i=90^\deg$ is edge-on), 
and the spin parameter of the black
hole, $a_\ast = J c/G M^2$, where $J$ is the hole's angular momentum.
If the hole is rotating, we assume that the disk lies in the
equatorial plane of the black hole, as a result of the
Bardeen--Peterson (1975) alignment mechanism, and that the disk is
corotating.  Of these parameters, perhaps the most problematic is the
disk emissivity.  The distribution of the hard X-ray flux which
illuminates the disk determines the emissivity of the
fluorescing disk material, and therefore has a strong influence on
line profiles.  The emissivity is usually assumed to be
axisymmetric.
This is reasonable if the observed profile is obtained with a long
integration time since strong asymmetries presumably would average
out.
Specific choices for emissivity include a power-law in radius (Fabian
et al. 1989; Bromley, Chen~\& Miller 1997), a form consistent with a
point source of illumination (Matt, Fabian~\& Ross 1993; Matt,
Fabian~\& Ross 1996; Reynolds \& Begelman 1997), and a function
proportional to the total energy flux in the Page \& Thorne (1974)
accretion model (Dabrowski et al. 1997). Dabrowski et al. (1997) also
considered a non-parametric form of the emissivity function. 
Recently, Iwasawa et al. (1999) performed a line-profile fit
assuming nonaxisymmtric X-ray illumination of the disk in MCG--6-30-15.
Nonaxisymmetric emissivity might be expected for the bright state
of this source if the mass of the central black hole
is $\sim 10^8\,M_{\odot}$. 
In this case the integration time of the observations
is shorter than the orbital period at a radius of a few times
that of the horizon.

The calculation of emissivity may be complicated somewhat by the local
physics of the disk as well as the nature of the illuminating
source. If the incident radiation is strong it can cause iron to
become highly or fully ionized. 
For iron atoms at ionization stages
no higher than Fe~\uppercase{xvi}, K$\alpha$ emission
occurs at 6.4~keV, however, resonant absorption and successive Auger processes 
prevent significant line emission 
by Fe~\uppercase{xvii} through Fe~\uppercase{xxiii}. 
Lithium-- and Helium--like 
ions, Fe~\uppercase{xxiv} and Fe~\uppercase{xxv}, emit at approximately
6.7~keV, while Hydrogen--like ion Fe~\uppercase{xxvi} produce 
K$\alpha$ line at 6.97~keV (Matt et al.~1996; Matt et al.~1993; see also the 
review by Fabian et al. 2000).

Local anisotropy of rest-frame emission is another effect which can
influence an observed line profile. It is often assumed that the
emitter is locally isotropic, although Laor~(1991) considered limb
darkening and Matt et al.~(1996) considered effects of resonant
absorption and scattering which can result in anisotropic emission.

The next level of detail in modeling local disk physics was to
consider non-Keplerian flows, as when material falls within the
innermost stable orbit around a Schwarzschild black hole (Reynolds \&
Begelman 1997), or when there is turbulence in a disk of finite
thickness (Pariev \& Bromley 1998).  More recently, temporal
variations in line profiles (Lee et al., 2000; Iwasawa et al., 1999;
Nandra et al., 1999) have been observed. Simulations of line profiles
with a time-dependent illumination (Reynolds et al. 1999; Young \&
Reynolds, 2000; Ruszkowski 1999) are also performed.  Knowledge of the
line variability can provide information about the size of the inner
region of the accretion disk and the mass of a black hole.

To circumvent uncertainties in the emissivity of the disk 
and yet still obtain estimates and constraints of fundamental disk
parameters such as the inner radius, Bromley, Miller \& Pariev (1998)
suggested using the minimum and maximum frequency shifts of a broad
emission line as a diagnostic. As first demonstrated by Cunningham
(1975), the edges of a line profile from an emitting disk annulus
uniquely determine the radius of the annulus and its
inclination. Similarly, bounds may placed on these parameters when the
emission comes from many such annuli.  These bounds are more robust
indicators than estimates derived by fitting the full profile shape
when the emissivity is nonaxisymmetric or when the continuum has been
poorly modeled in the central regions of the broad lines.

In this work we use the method of line frequency extrema to study the
geometric and kinematic properties of relativistic accretion disks.
We provide new methods of line edge detection based on polynomial fits
to the line profile and continuum, and compare the results to the $\chi^2$
rejection technique used by Bromley, Miller \& Pariev (1998).  
In \S~2 we describe the minimum and maximum frequency
diagnostic and ways to determine the position of a line edge and its
error. In \S~3 we apply the technique to the iron line profiles from
MCG-6-30-15, NGC~4151, and NGC~3516, and we compare the results of the edge
fitting to the fitting of the whole line profile. 
Finally, we discuss prospects for our method with data from
future X-ray missions.

\section{The Method of Minimum and Maximum Frequency Shifts}

\subsection{Description of the Method}

The information contained in a line profile is rich, but the parameter
space of models is also fairly large.  Here, as in Bromley, Miller \&
Pariev (1998), we sacrifice detailed modeling for broad characteristics
of the accretion disk by examining only the edges of the line. The
gain is a reduced dependence on unrealistic model assumptions such as
axisymmetric emissivity or the behavior of the continuum emission over
the entire frequency range of the line. For example, the usual
continuum model is a power law in frequency, but there is evidence
from Iwasawa et al.~(1996) that the index can vary in time.  Since the
lines are extraordinarily broad, one risks the possibility of a poor
guess for the continuum which would affect the shape of the inferred
profile. This in turn can affect inferences of the disk model
parameters, including inner and outer radii. Worse, the profile might
also be misinterpreted because our assumptions of axisymmetry might be
incorrect, a possibility which is worrisome in light of time
variability of the continuum and line profile. For example, patchy,
local flares would violate axisymmetry unless the observed line
profile were integrated over a long period of time.

From Cunningham (1975), it is clear that minimum and maximum 
frequency shifts of a line profile
themselves contain information about fundamental disk parameters,
regardless of details which may complicate the shape of the profile.
Here we use the frequency extrema diagnostic for
both standard infinitesimally thin disk case and for the disk model
taking into account turbulent broadening and additional frequency
shifts of the line due to Doppler effect associated with the accretion
inflow and correction to the gravitational redshift because of the
elevation of the line emitting spots above the equatorial plane of the
disk. For the latter case we use results by Pariev~\& Bromley (1998)
for the line profiles emitted from the surface of the ``standard''
$\alpha$-disk (Novikov \& Thorne 1973). 
The parameters which uniquely 
define the system of black hole and accretion disk are the mass of
the black hole $M$, spin parameter of the black hole $a_{\ast}$, and
inclination of the disk relative to the observer $i$. 
The shape of a 
time-independent line profile cannot provide us with
the absolute scale of the system, $GM/c^2$, although
variable features in the profile can indicate the
black hole mass. 
In the present work we deal only with time-averaged observations,
hence the mass $M$ enters only as degenerate scale factor with disk
radius.  In the case of the Shakura~\& Sunyaev (1973) $\alpha$-disk
model, additional model parameters include the luminosity of the disk
in Eddington units, $L/L_{\rm ed}$, and a viscosity parameter
$\alpha$. Of course, with fixed values for these parameters, an
observation of the luminosity can constrain the black hole mass.

Here we first define $g_{\rm min}=\nu_{\rm min}/\nu_e$ and $g_{\rm
max}=\nu_{\rm max}/\nu_e$ to be the minimum and maximum frequency
shifts of a line relative to the rest-frame frequency.  Then we
consider emission from an infinitesimal annulus of the disk with
Boyer-Lindquist radius $r$.  For a given value of the spin parameter
$a_{\ast}$ (as well as luminosity $L$ and viscosity parameter
$\alpha$ for a thick disk) there exists a unique mapping $g_{\rm
min}=g_{\rm min}(r,i)$, and $g_{\rm max}=g_{\rm max}(r,i)$,
connecting frequency extrema 
of a line and the inclination angle $i$ and radius
$r$ of the annulus. An example of such a mapping is given in
Figure~\ref{fig1}.  Note that we use units of gravitational radius,
$R_g=GM/c^2$, so that for a nonrotating black hole the radius of event
horizon is $2 R_g$, and the radius of the innermost stable orbit is $6
R_g$; for the maximally rotating astrophysical black hole with
$a_{\ast}=0.998$ (Thorne 1974), the radius of the 
event horizon decreases
to $1.063 R_g$ and the accretion disk becomes stabilized down to the
radius $1.237 R_g$. It is this extension of the stable disk down to 
small radii which is the most prominent effect of rotation of the black
hole.
 
Emission in an observed line consists of the sum of contributions
from many infinitesimal annuli. These annuli produce a curve
in the $g_{\rm min}$--$g_{\rm max}$ plane, which corresponds to
a single, constant inclination angle $i$.
Of course a measured profile will generate a single point
in this plane which is simply the extreme values of
redshift and blueshift along a constant-$i$ curve.
The position of the measured point
in the $g_{\rm min}$--$g_{\rm
max}$ diagram indicates that emission can come only from the quadrant of the
$g_{\rm min}$--$g_{\rm max}$ plane  
defined by inequalities $g(i,r)>g_{\rm min}$ and $g(i,r)<g_{\rm max}$. 
At the same time, some emission must come from the points on both
sides of this quadrant, i.e. from the points having either 
$g(i,r)=g_{\rm min}$ or $g(i,r)=g_{\rm max}$.
The position of this point can 
fix the range of possible disk
inclination angles $i$:
Since $g_{\rm min}$ increases with increasing $r$ along the curve
$i=\mbox{constant}$, the upper limit on inclination angle of the disk
is given by the value of $i$ for the curve $i=\mbox{constant}$ passing
through the data point.  The lower limit is provided by the
$i=\mbox{constant}$ curve which just touches the horizontal line
$g=g_{\rm max}$. If $g_{\rm max}<1$, then the lower limit for $i$ is
0. Furthermore, for each $i$ within the observed bounds one can
determine the outer $r_{\rm out}(i)$ and inner $r_{\rm in}(i)$ radii
of annulus, which give a contribution to the observed
profile. 
Generally, the middle regions of the disk can have
arbitrary amounts of iron line emission, but there must always be some
emission coming from points of the disk surface at radii 
$r=r_{\rm out}(i)$ and $r=r_{\rm in}(i)$.

If several line profiles are available for the same object in
different phases of emissivity, then we may be able to place further
constraints on the inclination angle of the disk and to check the validity
of the accretion disk model (i.e. that bounds for $i$ are not mutually
exclusive). Narrower limits for $i$  put tighter
constrains on $r_{\rm in}$ and $r_{\rm out}$.  We emphasize that these
constraints are independent of the emissivity law across the disk.

As mentioned above, by looking only at the position of the edges of
line profiles we clearly lose a large amount of the information
contained in the shape of a line profile. We instead obtain very
strong bounds on the inner radius and inclination angle of
the disk. In the case of time variability over time scales not
resolved during photon counts integration time, positions of the
frequency extrema are still able to provide estimates of the
inclination angle of the disk and bounds for the inner edge of the
disk.  This is an important feature when there are changes in the disk
illumination and/or possible obscuration of parts of the disk by
absorbers (Weaver \& Yaqoob 1998) which can substantially
alter the main core of the line.

Another advantage of looking only at the positions of
the edges of the line is that it circumvents the
effects on the line profile of possible resonant absorption of the
line photons in the disk corona. As shown by Ruszkowski \&
Fabian (1999), reasonable assumptions about the velocity field in
the corona of the disk suggest that an absorption feature is located
slightly redward of $6.4\,\mbox{keV}$ leaving the red and blue edges
of the line unaffected. In this case a
direct fit of the profile with a standard Keplerian disk model would
give poor results. The frequency extrema method does not require detailed
models of absorbing corona or occulting cloud, nor is
it sensitive to
the reflection from a surrounding torus or
any other sources of narrow $6.4\mbox{keV}$ iron line emission
except in the rare instances when the disk is observed almost
directly face-on and all disk emission
is reddened.

\subsection{Theoretical $g_{\rm min}$--$g_{\rm max}$ Maps}

We used a general purpose ray-tracing code to calculate values of
$g_{\rm min}$ and $g_{\rm max}$ for a number of narrow rings and 
a number of different inclinations $i$.
The code, described by
Bromley, Chen~\& Miller (1997),
generates a pixelized
image of the accretion disk as would be seen by a distant
observer. The observed frequency at each pixel is given by
\begin{equation}
   \label{eq:gee}
   g \equiv \frac{\nu_o}{\nu_e} = \frac{-1}{-\vec{u}\cdot\vec{p}}
\end{equation}
where subscripts $o$ and $e$ are observer and emitter respectively,
$\vec{u}$ is the 4-velocity of the emitter and $\vec{p}$ is the
emitted photon's 4-momentum.  Note that the emitter 4-velocity is
specified by the disk model, while the photon 4-momentum is calculated
by numerically tracing the photon geodesic back in time from the pixel
in the observer's sky plane to the surface of the disk. Then, we sort 
pixels into narrow rings and determine extrema of $g$ over all pixels 
fallen into a ring between $r$ and $r+\delta r$. To create a grid of 
lines $i=\mbox{constant}$ and $r=\mbox{constant}$ on 
$g_{\rm min}$--$g_{\rm max}$ plane we use cubic spline interpolation.

The ray tracer itself is a general-purpose 
second-order geodesic solver for a Kerr geometry. An arbitrary
disk surface can be specified so that the photon trajectories
terminate on this surface. For axisymmetric geometries the number of 
individual geodesics which need to be traced numerically in order 
to produce complete set of images
is of the same order as in the method of transfer 
functions by Laor (1991). Using of parallel supercomputer allowed us
to compute an image of 1000$\times$1000 pixels during about 1 minute 
of run-time.

Details about modeling the emitter
4-velocity, $\vec{u}$ in equation~(\ref{eq:gee}), can be found in
Pariev~\& Bromley (1998). Here, we assume that outside of the orbit of
marginal stability the bulk emitter 4-velocity is the sum of 
a Keplerian 4-velocity and a small, inward radial velocity component as
given by Novikov~\& Thorne (1973). We do not take into account the $\theta$
component of inflow velocity, nor the dependence of 
$\phi$ component of velocity on the height of the disk. 
These are higher order corrections to radially directed inflow.  
Inside of the marginal stability radius, all
orbits are presumed to be in free-fall in the equatorial plane 
with integrals of motion equal to the values at the innermost stable 
orbit.  
We assume a simple model for turbulent motion in the 
disk, with an isotropic Gaussian distribution of turbulent velocities 
with the square mean equal to the square of the speed of sound 
$c_s^2$ averaged over the disk thickness, namely 
\begin{equation}
I(\nu_e,\nu,\mu_e,r_e)=\epsilon(r_e,\mu_e)\sqrt{\frac{3}{2\pi}}\frac{c}{c_s}
\exp\left[-\frac{3c^2}{2c_s^2}\frac{(\nu-\nu_e)^2}{\nu_e^2}
\right]\frac{1}{\nu_e}\label{broad}\mbox{,}
\end{equation}
where the intensity $I$ at a specified frequency $\nu$ depends upon
the rest frame energy of the $K\alpha$ line ($\nu_e=6.4\,\mbox{keV}$),
the angle cosine $\mu_e$ of the photon emission with respect to the
normal of the disk as measured in the source frame, and the radial
coordinate $r_e$ of the emitting material on the surface of the disk;
$\epsilon(r_e,\mu_e)$ is the surface emissivity; $c_s(r_e)$ is the
sound speed at the radius of emission,~$r_e$. We consider only 
the case of isotropic emission, when $\epsilon=\epsilon(r_e)$.
The width of a Gaussian line profile in the comoving frame with 
the disk surface is $\sigma(r_e)=c_s(r_e)/\sqrt{3}$.
The sound speed at a radius $r_e$ is proportional to
the accretion rate and, therefore, to the luminosity of the disk
(Shakura~\& Sunyaev 1973; Novikov~\& Thorne 1973; equation~[15] 
in Pariev~\& Bromley 1998). Therefore, the amount of smearing 
of the line at a given radius is proportional to $L/L_{\rm ed}$.
Note, that the actual turbulent velocities may have a mean value less 
than the speed of sound (see Pariev~\& Bromley~1998 for details), thus
we are considering the maximum possible effect of turbulence on line profiles.

Turbulent broadening causes the whole line profile to be smoothed.
Particularly, blue and red edges are no longer sharp but have wings 
of the order of $c_s/c$ at the radii where the edges of the profile are
formed. Accretion inflow causes a slight shift of the profile to the 
red, but does not strongly influence the position of the profile edges,
since the edges are formed in regions of the disk which have radial inflow 
velocities nearly perpendicular to the light ray emerging toward the 
position of the observer.

Adding a Gaussian turbulent velocity spectrum makes the
determination of edges of a theoretical line profile somewhat
uncertain. Moreover, the observed extent of a line depends upon the
signal--to--noise ratio of the observational data.
Keeping in mind that our Gaussian
prescription for turbulent broadening is crude and
velocities of turbulent motions along line of sight cannot exceed the
speed of sound, we assume the following procedure for finding the
minimum and maximum redshifts in the case of a turbulent disk model
with finite thickness and accretion inflow: We 
add the value of $\sqrt{3}\sigma(r)/c=c_s(r)/c$ to the uncorrected 
$g_{max}$ in
order to obtain $g_{\rm max}$ and subtract the value of $c_s(r)/c$
from the uncorrected $g_{\rm min}$ in order to obtain $g_{\rm min}$. 
This procedure is also consistent with a signal--to--noise ratio
of about 10\% in currently available observational data for line profiles.
The grid of $g_{\rm min}$--$g_{\rm max}$ obtained in this way was used
in all estimates of parameters of thick disk model and is shown
by dashed lines in Figs.~\ref{fig1} and~\ref{fig5} for nonrotating and extremely 
($a_{\ast}=0.998$) rotating black holes. To determine the greatest
possible effect of 
the physical structure of the disk, we have chosen a high
disk luminosity, $L=L_{\rm ed}$, and an $\alpha$ viscosity
parameter of $0.3$. These values still consistent with an optically thick
disk. 

The main effect kinematics and finite disk thickness
is to yield larger values for $g_{\rm max}$ and smaller
values for $g_{\rm min}$.  For a nonrotating black hole the disk extends
down to $6 R_g$, at which point the
thickness of the disk as well as radial inflow velocity
approach zero, while the surface density increases without bound (Novikov \&
Thorne 1973) --- this is a consequence of the zero-torque boundary condition
at the inner edge of the disk. Thus 
the difference between the thin and thick disk models vanishes at
$6 R_g$. In reality, the gas smoothly transits near $6 R_g$ from a slow
inward spiral caused by viscous stresses to a plunging geodesic 
orbit into the black hole. 
Nonzero viscous torque and radial inflow should also
change the parameters of the disk just beyond the $6 R_g$ orbit
(e.g. Beloborodov, Abramowicz~\& Novikov, 1997).  Thus, the shape of
the dashed curves in Fig.~\ref{fig1} for $r\le 10 R_g$ should be altered
for more realistic situations so that the transition from
$r>6 R_g$ to $r< 6 R_g$ is smoothed out and dashed lines should
move upward and to the left near the $r=6 R_g$ curve of the thin disk
model. 

Agol \& Krolik~(2000) consider thin-disk structure when there
is a finite time-steady torque on the inner edge of the disk. 
According to them, additional dissipation near innermost edge of 
the disk due to the applied torque would cause increased heating
and increased sound speed compared to Novikov-Thorne disk. Larger
sound speed causes dashed lines on Fig.~\ref{fig1} to move 
upward and to the left. However, Paczynski~(2000) points out 
that any torque at the innermost stable circular orbit should
be small as soon as the disk is thin and $\alpha\ll 1$. The subject
of the transition from nearly Keplerian flow to rapid infall to
the black hole is under active investigation (e.g., Gammie 1999). 
We only mention that our method of line frequency extrema diagnostic could be 
used to put constrains on models resulting from such investigations,
but this is beyond the scope of the present work. 

The deviation of frequency extrema plots for disks of finite
thickness from those for
infinitely thin disks scales as $\propto L/L_{\rm ed}$ with the total
thermal luminosity of the disk $L$, while the exact value of the $\alpha$
parameter is not nearly as important (Pariev~\&
Bromley, 1998). 
When then accretion rate and the ratio $L/L_{\rm ed}$ decrease, the
sound speed, radial inflow velocity and thickness of the disk also
decrease. In the limit of very small accretion rates and very small
$L/L_{\rm ed}$, the solid and dashed grids in Fig.~\ref{fig1} and
Fig.~\ref{fig5} becomes coincident. 
For very low accretion rates the
disk can be regarded as thin, with each light-emitting element moving on
a circular Keplerian trajectory with little turbulent broadening.

\subsection{Line Blending}

Blending of emission lines having different rest-frame frequencies 
would influence results of using frequency extrema method. The most 
important sources of
blending can be the presence of K$\alpha$ emission from
highly ionized iron, emission of iron K$\beta$ at $7.06\,\mbox{keV}$, and 
nickel K$\alpha$ $7.48\,\mbox{keV}$ line (George \& Fabian 1991).
The most difficult to estimate is the contribution
of the hot (i.e., highly ionized) iron K$\alpha$ line.
Hot disks can fluoresce
in the K$\alpha$ line under certain circumstances, depending on the
three-dimensional temperature structure of the disk.  These issues
were given consideration by Matt et al.~(1996), Matt et al.~(1993), 
and \.{Z}ycki~\& Czerny~(1994): The basic
conclusion reached by these authors is that the rest-frame frequency,
intensity, and angular dependence of the emission are determined by the
value of the ionization parameter $\xi(r)=4\pi F_X(r)/n_{\rm H}$, where
$F_X(r)$ is the X-ray illuminating power-law flux striking a unit area
of the disk surface, and $n_{\rm H}$ is a comoving hydrogen number
density. In order to
determine $\xi$ one must know the characteristics of the source of
illuminating X-rays such as its intensity, spatial distribution, and
motion relative to the disk. Some aspects of the dependence of the
line profile and equivalent width on these characteristics have been
outlined in Reynolds~\& Begelman~(1997) and Reynolds~\&
Fabian~(1997). Our estimate of the parameter $\xi$ along the lines
described in Reynolds~\& Begelman~(1997) shows that for a
 thin $\alpha$--disk around a Schwarzschild black hole
 it can plausibly lie either above or
below the threshold value for ionization
$200\,\mbox{erg}\,\mbox{cm}\,\mbox{s}^{-1}$, depending upon X-ray
efficiency of the illuminating source and its spectral index. For the Kerr
case when the disk extends close to the event horizon, relativistic
aberrations of the illuminating radiation due to gravity, frame dragging,
and the motion of material in the disk become very significant, generally
leading to the enhancement of the  irradiating flux as measured in the
comoving frame of the disk material.  Consequently, the parts of the
disk at small radii are more likely to emit hot iron lines $6.67\,\mbox{keV}$
and $6.97\,\mbox{keV}$. 
However, due to resonant scattering and Auger processes, little line 
emission is produced in the transition region from cold to hot iron
lines.

Furthermore, in regions where lines can form, X-ray flares above the disk 
can cause higher ionization of the disk surface directly below the flare
than the rest of the disk. All this makes the analysis of the iron line 
profile complicated and dependent upon unknown positions 
of illuminating X-ray sources, accretion rates and mass of the central
black hole. We assume that the iron K$\alpha$ line has $6.4\,\mbox{keV}$
rest frame energy everywhere in the accretion disk. For all observational
data considered here, the core of the line is centered close to 
$6.4\,\mbox{keV}$, while higher energies for the core can be ruled out with
very large degree of confidence. This means that parts of the disk
not very close to the black hole ($>20 GM/c^2$), where the core of the
line is formed, produce the cold iron $6.4\,\mbox{keV}$ line. One is not so 
certain about inner parts of the disk producing the red tail of the line,
but we assume that the K$\alpha$ line has the rest frame energy of
$6.4\,\mbox{keV}$ throughout the disk. The maximum possible error in finding 
$g_{\rm min}$ introduced by this assumption is $0.09$, which in most 
cases is no more than the size of $2\sigma$ error box for the values of
$g_{\rm min}$ for the present quality of X-ray observations. 

Since the K$\beta$ $7.06\,\mbox{keV}$ line is emitted by the same iron
atoms that are responsible for the K$\alpha$ line, the shape of 
the two profiles should be the same.
The ratio of fluorescent yields in 
K$\beta$ and K$\alpha$ lines is $0.113$ (George \& Fabian 1991).
If one makes the reasonable assumption that 
nickel is distributed over the disk surface in the same way as
iron, then the profile of nickel K$\alpha$ line will be similar to
that of iron, but shifted by $1.08\,\mbox{keV}$. The 
relative abundance of nickel with respect to iron is not known precisely 
for observed
sources but a reasonable value is about $0.06$ 
(George \& Fabian 1991). 

The fact that the iron K$\beta$ line and the nickel K$\alpha$ line 
are faint compared to iron K$\alpha$ allows us 
to perform an effective decomposition
of the blended profile into the sum of individual profiles of three 
lines. If $f_{\rm obs}(\nu)$ is the observed excess of photon counts per
second over the fitted continuum, and $f(\nu)$ is the actual profile of iron 
K$\alpha$ line, then one has the relation
$$
f_{\rm obs}(\nu)=f(\nu)+0.113 f(\nu-0.64)+0.06 f(\nu-1.08)\mbox{.}
$$
Here we will use the term ``energy-corrected profile'' for the 
function $f$.
One can solve this equation for $f$ iteratively, feeding
approximations to the function for the contributions from
the K$\beta$ and nickel lines. As an
initial guess, we take $f_{\rm obs}$ and each 
subsequent approximation $f_{k+1}$ is obtained using
previous values of $f_k$:
$$
f_{k+1}(\nu)=f_{\rm obs}(\nu)-0.113 f_k(\nu-0.64)-0.06 f_k(\nu-1.08)\mbox{.}
$$
The actual data points are the integrals of a continuous energy 
distribution within 
energy bins. The accuracy in determining $f(\nu)$, which we need according
to the quality of the data, is of the order of a few per cent. Therefore, 
it is sufficient for our purposes to treat data points just as a discrete 
samples of the underlying continuous function.
We computed values of $f_{k+1}(\nu)$ at the center of each energy bin,
linearly interpolating between bins to obtain values of  
$f_k(\nu-0.64)$ and $f_k(\nu-1.08)$.
When the shifted frequency went beyond the red boundary of the
observed data, we used a value of zero for the
shifted profile. For the data considered here,
the iteration procedure converged with sufficient accuracy in
only two iterations.

The main effect of the blending correction is to suppress the blue
wing of the line or perhaps create an ``absorption'' feature if the
blue wing is weak or absent.  We compared results of determining
$g_{\rm max}$ using uncorrected and energy-corrected profiles for the
same observation, we found that only in the data with the highest
signal-to-noise was there a difference in the detected edge position
of greater than $2\sigma$. Thus, at the quality of present X-ray
spectral data for most Seyfert galaxies, the corrections for blending
with lines other than iron K$\alpha$ line are small.

\subsection{Edge Detection}

We are interested in the determining boundary between the line
and continuum, not the full profile shape, and we now argue that we can
work with the observed counts directly without restoring actual X-ray
fluxes by deconvolving with the response matrix of the X-ray detector.
All the data considered below were obtained with the ASCA SIS
detector; its sensitivity to a monochromatic line varies by a factor of five
over the 3-8~keV waveband, with the response function broadening the
signal by about 100~eV.  One or two satellite peaks also appear at
a level not exceeding 3\% of the intensity of the main line. The
frequency of the satellite peak is about two thirds that of the
initial line. In addition, the monochromatic line produces a small
continuous response in each channel with counts less than 1\% of the
counts in the channel at the peak.  The magnitudes of detector
response effects are well within the errors of the observed intensity
of the line 
and the spread over 100~eV is comparable to
smallest standard error in the position of the edges measured
here (see Tables~1 and~2 with the results for edges). Thus, we
conclude that folding the broad line and continuum spectrum through
the detector response matrix does not have significant (i.e. more than
one standard error) influence on the locations of the line edges.

To locate the edge of a line profile, we start with an unconvolved,
continuum-subtracted, energy-corrected line profile as observed in the
restframe of the host galaxy (i.e., cosmological redshift has been
removed). 
The continuum subtraction was performed differently
by different observers: 
Iwasawa et al.~(1996) performed a fit with a single power law absorbed
by the Galactic column density 
($N_H \approx 4\times 10^{20}\,\mbox{cm}^{-2}$) of the 
underlying continuum in 3-10~keV range for MCG-6-30-15. Iwasawa 
et al.~(1999) slightly improved the continuum fit by adding a reflection 
component modeled by {\tt pexrav} task (Magdziarz~\& Zdziarski 1995).
However, this reflection component affects line flux by only 
about 5 per cent. Wang et al.~(1999) fitted the continuum in the 1.0--4.0
and 8.0--10.0~keV band with a model which consists of a dual absorbed
power law with some fraction (about 5 per cent) of the direct continuum 
scattered into our line of sight and absorbed only by the Galactic
column. Compton reflection was not included in this fit since there
is no reliable indication of the existence of the reflection in 
this object. Nandra et al.~(1999) fitted the continuum in the 
3.0--4.0 and 7.0--10.0~keV range by a single power law only.
  
We need to examine only sections of the profile containing a
suspected line edge and fit the profile with a linear, polynomial
model $y=a_0+a_1 x + a_2 x^2 + a_3 x^3$ in frequency $x$.  In most
instances, a quadratic fit gives the best results, since the cubic
term added artificial oscillations to the fit. A linear $\chi^2$
method determines the polynomial coefficients $a_i$, and these in turn
provide roots of the polynomial model.  The real root (if it exists)
which is closest to the center of the line is taken as the position of
the edge.

To estimate the error in
measuring the energy of the edge of the line using the roots of a
best-fit polynomial we use Monte-Carlo realizations of the 
observed counts:
Each profile consists of a number of counts $y_i$ in energy
channels spanning the intervals from $x_i-\Delta_i$ to $x_i+\Delta_i$,
where $x_i$ are the central energies of each channel number $i$, 
$\Delta_i$ are half--widths of the channels. The half--widths of
the channels increase with increasing energy and range from
about 50~eV at 3~keV to about 700~eV at 9~keV. 
For each real observed profile we create 20000 artificial data sets by
randomly drawing a flux value $y$ from a Gaussian
distribution with mean $y_i$ and variance equal to the width of
$y$-error bar $\sigma_i$. Then we perform a $\chi^2$ minimization of
each artificial data set to get the best-fit polynomials and
their roots.  Repeating this procedure for all 20000 Monte-Carlo data
sets allows us to construct a probability distribution for the
location of the line edges determined in this way. Generally, such
distributions are skewed with tails extending out from the center of
the line.  We calculate the mean and the variance of this distribution
of edges. Then, we take the mean as the best approximation for the 
true position of the edge and variance as a standard error of this 
edge determination.
To be more conservative in light
of the fact that the distribution of best-fit values is not strictly Gaussian,
we take our final values of $g_{\rm min}$ and $g_{\rm max}$ to have
error boxes twice as large as the standard error given by Monte-Carlo
distribution of fits.  In most cases, the ellipse in the $g_{\rm
min}$-$g_{\rm max}$ plane with semimajor axes that are twice
the standard errors contain 95\% of all roots found.

The results of fitting with the polynomial model are given in Table~1.
For the fits we choose only channels on one side of the line
peak, either red or blue depending on which edge we are looking for.
Each set of channels are continuous in energy and define some
interval which brackets a line edge, i.e.,
the interval contains regions where the spectrum is well-fit
to the continuum model as well as channels which exhibit
clear and significant signal from line emission.
Once chosen, the set of channels remains the same for all fits in
a Monte-Carlo simulation. The actual energy range covered by the fitting
intervals varies substantially for different profiles depending on
how sharp is the transition from the line to continuum. Roughly,
energy range is about 2--3~keV for the red edge and 1--2~keV for the 
blue edge. Instead of specifying the energy range for each fit
we choose to number all points on the plot of the profile and to
list those which are used in a particular fit.   
The points are numbered consecutively with the point number~1 referring
to the redmost channel. The numbers of points used in the fits are 
given in the third column of Table~1. 

We also propose a non-linear ``sharp edge'' model to locate a line edge.
In this model the continuum
is exactly zero and the line is a linear or quadratic piece which
extends from the continuum to higher intensities.
At the red edge of a line the model is
\begin{eqnarray}
&& y=0 \quad \mbox{for} \quad x<p_0\mbox{,} \nonumber\\
&& y=p_1(x-p_0)+p_2(x-p_0)^2 \quad \mbox{for} 
\quad x>p_0\label{fitred}\mbox{,} 
\end{eqnarray}
while for the blue edge
\begin{eqnarray}
&& y=0 \quad \mbox{for} \quad x>p_0\mbox{,} \nonumber\\
&& y=p_1(x-p_0)+p_2(x-p_0)^2 \quad \mbox{for} \quad x<p_0\label{fitblue}\mbox{,} 
\end{eqnarray}
where in the last case $p_1$ is negative. The
line edge is $p_0$, the location of the kink which delineates the
flat continuum and the regions of increasing flux $p_0$.

As with the polynomial fitting, we determine the distribution of $p_0$
in many Monte-Carlo realizations, and take the mean and variance of
this distribution as the estimated position of the edge and 
its corresponding uncertainty.
The fitting function (\ref{fitred})-(\ref{fitblue}) is not
smooth and this causes many fits to give $p_0$ values very close to a
mean channel frequency $x_i$, mimicking a discrete distribution.
Therefore, the cumulative distribution of $p_0(x)$ looks like the sum
of several step functions.  The contour around the mean point $g_{\rm
min}$--$g_{\rm max}$ determined from the Monte Carlo simulations which
characterizes the scatter in the $g_{\rm min}$--$g_{\rm max}$
simulation points is rectangular in shape rather than the usual
ellipse (as in the case of Gaussian distributed $g_{\rm min}$ and
$g_{\rm max}$). We verified that for most fits the rectangle, centered
on the mean value for the Monte-Carlo simulated $g_{\rm min}$ and
$g_{\rm max}$ values and having sides equal to twice the variance of
the Monte-Carlo simulations, contains about 95\% of fitting points
inside it.  We plot these $2\sigma$ ``error-rectangles'' as well as
best fit to actual data points on the $g_{\rm min}$--$g_{\rm max}$
diagrams in Figs.~\ref{fig5}--\ref{fig7}.
In order to make Figs.~\ref{fig6}--\ref{fig7} easier to read,
we use asymmetric error bars instead of full ``error rectangles.''
These error bars originate at the point corresponding to the best 
fit to the actual data. The ends of all four error bar segments define 
the error rectangle, i.e. the length of each of the 
four error bars is equal to the distance between best fit point 
and an edge of the rectangle. All our conclusions about 
disk geometry assume that the error distributions lie within such 
rectangles.   
Table~2 shows edge estimates
based on this non-linear model, analogous to Table~1.

In order to verify that the non-linear model gives reasonable determinations
of the positions of the line edges we applied the same Monte-Carlo
method for finding the edge of an artificial line profile.
This mock line profile was created from a real profile 
in the following way. Starting with the same energy channels $x_i \pm 
\Delta_i$ as for the real data set, we find the best fit of the
non-linear model to the actual data set $y(x)$ and take values 
of $y(x_i)$ of this fit. 
Then, using error values corresponding to the real data we
apply the same Monte-Carlo procedure of finding the edge 
of the artificial data set and thus determine the distribution 
of the parameter $p_0$. 
Generally, we obtain the
smaller errors and lower $\chi^2$ for this type of fit, while the 
one standard error intervals for $p_0$ found by both Monte-Carlo 
simulations are overlapping (see few exceptions for poor quality edges 
below). This result 
is expected since the most likely profile for the data is the
actual measured points, while the most likely profile for any
given Monte Carlo realization is the underlying model profile.

Which model, polynomial or non-linear, better approximates the true shape
of observed profile?
Theoretically one expects that the profile formed by 
a thin Keplerian disk has sharp edges.  
Let us consider the part of the disk surface near the innermost 
edge of the disk, where 
the reddest photons near the red edge of the line with frequencies 
from $g_{\rm min}$ to $g_{\rm min}+\Delta g$ are emitted. This part 
of the disk has the shape of the crescent elongated along the inner
edge of the disk (e.g., Fig.~8 in Luminet 1979, or color figures
in Bromley et al. 1997; Pariev~\& Bromley 1998).  
It is easy to see that the area of this crescent is $\propto \Delta g^{3/2}$. 
Therefore, the flux (or photon counts) per unit frequency near the red edge
of the line should be $F_\nu \propto \Delta g^{1/2}$ (if the illumination
intensity does not approach zero at the inner edge of the disk). 
Certainly, the coefficient in these proportionalities gets very small for
rapidly rotating black holes because of the large gravitational 
redshift close to the event horizon and strong relativistic decrease of 
the intensity of the emitted X-rays. The whole area of the disk, where 
the red edge of the line is formed ($\sim R_g^2$), is also small.   
The part of the disk forming the bluest photons has an elliptic shape 
centered at a point
far from the event horizon. Similar consideration for the 
blue edge of the line gives $F_\nu \propto \Delta g$. However, the 
coefficient here is large, because the region on the disk surface producing
the bluest photons is relatively large.
It is for this reason that observed line profiles show
much sharper, well-defined blue edges.
Sharp red edges can be noticed only for 
slowly rotating black holes but become less prominent
for rotating black holes. Indeed, line profiles for a 
Schwarzschild system, calculated 
with high resolution, have a much more distinctive 
sharp red edge (e.g., Fig.~5 in Pariev~\& 
Bromley 1998) than in the case of an extreme Kerr metric.
(e.g., Fig.~6 in Pariev~\& Bromley 1998). 

While polynomial fitting does not take into account the theoretically 
expected shape of the edge profile,
the nonlinear model (\ref{fitred})-(\ref{fitblue}) is an approximation to 
the real edge. 
However, many observed profiles do not show easily discernible sharp edges
in the presence of noise. 
In particular, red edges have the appearance of smooth transits between
line and continuum. Blue 
edges are sharper, but are often accompanied by a trough at higher
frequencies. Hence even in this case, the blue edge is
often better described by the polynomial model than the
non-linear model with its zero-continuum requirement. As a result
the latter is seen to have large errors in most cases.
However, when blue edges occur without a significant negative trough, the
nonlinear model gives a better fit than the polynomial profile. 

We have demonstrated through Monte Carlo experiments that
we can obtain reasonable values for line edges and get good 
estimates of uncertainties. We checked that our technique
is not sensitive to the functional form
of the line profile, a test which indicates how robust
our method is to the emissivity function of the disk and
the energy-dependent detector response.
Specifically, we verified that our fitting technique gives values of
$g_{\rm min}$ and $g_{\rm max}$ that are not overly sensitive to the
multiplication of the data by a smoothly varying, positive definite 
function.  This procedure resulted in substantial changes
in the variance of estimated roots. However, if the function does
not change greatly in the vicinity of the edge, the new mean of
Monte-Carlo simulations remained within the largest of the one
$\sigma$-errors for the non-modified and
modified data sets. 

We also checked our method using
two functions, one on either side of the best fit line edge, each
giving a multiplicative factor which modifies
the observed intensity,
to vary both the line and the continuum independently.
By choosing particular
shapes for these function one can either augment or suppress the line with
respect to the continuum data adjacent to the line. 
This mimics the effect of 
increasing or decreasing the emissivity of that part of the disk
which contributes to the line edge.  It
turns out that multiplication by such a function can significantly
change the error in the determination of the edge but the location of the
edge itself still remains within the largest of the one $\sigma$ errors for
the original and modified data.  In the case where the line is suppressed
and the tail of the line became very shallow, the error in the
position of the edge increases. Conversely, highlighting
the line and suppressing fluctuations in the continuum decreases the error.

These tests indicate that the Monte-Carlo method gives
robust estimates of the position of the edge without
great sensitivity to the details of the disk emissivity function.
Certainly, there are limits. For example, if the disk is
illuminated in an inhomogeneous fashion as a result of flares located
close to the disk surface, the line contribution from a highly
localized region of the disk  could be
mistaken for the red edge of the whole profile. In this case one would 
miss a faint red tail extending beyond this jump and get an
overestimate of the inner extension of the accretion disk.

As discussed below, our Monte Carlo procedure worked poorly for estimating
line edges in some particularly noisy data sets such as those of Iwasawa
et al.~(1996) for the Dark Minimum (DM) and Bright Flare (BF) states
of MCG-6-30-15.
The best fit for the red edge in DM data
gives negative values for $p_0$, while the Monte-Carlo scattering of
points around best fit for the red edge in BF data gives negative mean
value of $p_0$ in sharp contrast with the best fit of
$4.13\,\mbox{keV}$ to the actual data points.  To evaluate the
position of the red edge in these noisy data sets we can formulate
alternative methods for line edge detection which are based more
directly on statistical properties of the line. For example, given a
set of flux measurements as a function of energy, we could test the
hypothesis that the fluxes are all consistent with continuum emission,
using $\chi^2$ to measure of goodness-of-fit (e.g., Bromley, Miller \&
Pariev 1998).  An edge detection algorithm then might be to test the
continuum hypothesis on the red-most channels, including successively
bluer channels until the continuum hypothesis can be rejected at some
specified level of confidence. The edge of the line would then
identified with the energy of the bluest channel.

In principle, this ``running $\chi^2$'' method is conservative. It
requires only that enough channels be sufficiently different from the
continuum that we can reject the continuum model. If the continuum
were extremely well sampled, then possibly a sequence of outlying
fluxes all packed at, say, the blue end of the set of channels could be
dismissed as merely expected statistical fluctuations. A less
conservative method should consider correlations between the fluxes in
neighboring channels to enable a more sensitive edge detection.

A further drawback of the running $\chi^2$ method is that it cannot
distinguish between absorption and emission features.  In practice,
one can replace the flux in suspected absorption regions with exact
continuum values or even randomly generated values.  To down weight
the effects of possible absorption features we also consider a "sign"
statistic to estimate the likelihood that a set of fluxes contains a
line edge. Specifically, we count the number of fluxes above and below
the continuum and compare with the expected binomial distribution.

We applied both these methods to the line profile data and obtained
limits on the red edge of the lines in the Iwasawa et al.~(1996) Deep
Minimum and Bright Flare data. The two methods yielded similar results
(Figure~\ref{fig6}), with the edge detected by the sign method being
10\% bluer than that detected by the running $\chi^2$ method.

\section{Implications for MCG-6-30-15, NGC~4151, and NGC~3516 line profiles}

\subsection{MCG-6-30-15} 

MCG-6-30-15 is a nearby ($z=0.008$) Seyfert~1 galaxy. The broad skewed 
iron line profile observed from this galaxy has drawn much attention 
from the X-ray community over the past few years after its discovery (Tanaka 
et al. 1995).
Figure~\ref{fig2} shows the results of polynomial fitting for frequency 
extrema from
different available observations of MCG-6-30-15. 
Figure~\ref{fig6}
corresponds to the 
same profiles but fitted with the nonlinear edge detection 
procedure. 
Comparison of the edge estimates using the two
different methods indicates that errors from the nonlinear 
fit are larger than errors from the polynomial fit
(except in the ``97b'' data from Iwasawa et al. 1999,  for which the nonlinear
fit's error box is smaller than the error ellipse of the polynomial fit). 
Best fits with
nonlinear method are, generally, very close to the Monte-Carlo mean values 
for the polynomial fits, with the exception of BF profile from Iwasawa 
et al.~(1996). However, as we mentioned above, the nonlinear fitting 
of the BF data worked poorly, and the running $\chi^2$ test gives somewhat
redder position of the edge than polynomial fitting. 

We reported briefly on the application of $g_{\rm min}$--$g_{\rm max}$
method in Bromley et al.~(1998). The results for $g_{\rm
min}$--$g_{\rm max}$ for Int, BF, and DM data of Iwasawa et
al.~(1996), reported in Bromley et al.~(1998), were very
conservative, i.e. the extent of both red and blue edges in Bromley et
al.~(1998) were probably underestimated.  Our new edge detection
procedure, which is designed to isolate an edge, not to find a
bounding value, gives smaller $g_{\rm min}$ and larger $g_{\rm
max}$. In contrast, the $\chi^2$ rejection method used by Bromley et
al.~(1998) does not take into account correlations in positions of
data points, thus it may miss extended emission edges which smoothly 
approach the continuum but at intensity levels which
are comparable to the errors of
observations. Our new fitting technique allows us to hone in on
such significant edges, if they exist, although the uncertainties can be
larger in this case. In choosing our new method over the 
$\chi^2$ rejection technique, we are opting for precision over accuracy.

The ``Int'' data set gives the most restrictive point in the $g_{\rm
min}$--$g_{\rm max}$ plane for determining the lower limit of the disk
inclination angle. Using $2\sigma$ error limits for the position of
that point results in the following absolute lower limits on $i$: in a
thin-disk Schwarzschild system $i>36^\deg$ (nonlinear fitting,
Fig.~\ref{fig6}a) 
or $i>38^\deg$ (polynomial fitting, Fig.~\ref{fig2}a);
in an extreme Kerr system with a thin disk, $i>34^\deg$ 
(nonlinear fitting, 
Fig.~\ref{fig6}c)
or $i>36^\deg$ (polynomial fitting, Fig.~\ref{fig2}c).

Incorporating all five available data sets, marked as
T,97,97b,Int, and BF, one can estimate a likely lower bound on the
inclination of a thin disk as $i>44\pm 6^\deg$ (nonlinear), $i>48\pm 5^\deg$
(polynomial) for a Schwarzschild system, and $i>42\pm 5^\deg$ (nonlinear),
$i>45\pm 5^\deg$ (polynomial) for extreme Kerr system.

These results contradict the estimates of $i$ obtained from the
numerous fittings of the full line profile to thin-disk models:
$30.2^{\deg +1.5}_{-2.7}$, $29.7^{\deg +2.9}_{-3.9}$ (Tanaka et al.,~1995,
Schwarzschild model), $26.8^{\deg +2.1}_{-1.0}$ (Tanaka et al.,~1995, Kerr
model), $29^{\deg +2.5}_{-3.2}$ (Dabrowski et al., 1997), 
$32^{\deg +2}_{-2}$
(Iwasawa et al.,~1999). The most probable explanation for the
discrepancy is that the blue edge of the profiles is not as sharp as
predicted by the thin-disk models.  The positions of the points in the
frequency extrema diagrams for turbulent disk models
(Figs.~\ref{fig2}b, \ref{fig2}d, \ref{fig6}b, and \ref{fig6}d)
indicate that the effects of finite disk thickness is to reduce the
lower limit on the inclination of the disk. Smoothed blue wings are
characteristic of turbulent disk profiles as well (Pariev~\& Bromley,
1998). Comparing Figs.~\ref{fig2}a and~\ref{fig2}b, \ref{fig6}a
and~\ref{fig6}b, one can see that the model of a thick disk around
nonrotating black hole with the luminosity of $L\approx 0.5 L_{\rm
ed}$ is enough to bring the lower bound on $i$ from the position of
the Int data point down to $30^\deg$. To reduce the most probable inclination
estimate from the five data points (T,97,97b,Int,BF) to $30^\deg$
requires higher accretion rates at a level of $L\approx L_{\rm
ed}$. Since the efficiency of converting energy to radiation is higher
for an extreme rotating black hole, the effect of turbulent motions is
smaller for the same values of $L$. By comparing Figs.~\ref{fig2}c
and~\ref{fig2}d, \ref{fig6}c and \ref{fig6}d, one can see that at
$L=L_{\rm ed}$ the absolute $2\sigma$ limit on $i$ reduces to
$30^\deg$, while one needs even higher accretion rates in order to
diminish the most probable lower bound on $i$ down to $30^\deg$.  We thus
conclude that the MCG-6-30-15 data suggest either a thin disk at
relatively high inclination angle or a thick disk at lower
inclination.

A key factor for our hypothesis that MCG-6-30-15 may harbor a thick
disk is the mass of the black hole, since this determines the ratio of
$L/L_{\rm ed}$.  There is no compelling measurement of the mass of
central black hole in MCG-6-30-15 to date. However, the form of the X-ray power
density spectrum reported in Nowak~\& Chiang (1999) evidently 
is universal, spanning low mass ($\sim 10\,M_{\odot}$) objects like of Cyg~X-1
to supermassive 
($10^8\,M_{\odot}$) ones like  NGC~5548.
The scaling of the frequencies at
which the power laws break led Nowak~\& Chiang (1999) to conclude
that the mass of the black hole in MCG-6-30-15 is about
$10^6\,M_{\odot}$, which is at the lower end of the supermassive black
holes mass spectrum in AGNs (Richstone et al., 1998). 
Reynolds~(2000) looked at possible reverberation delays 
between continuum emission in energy bands of 2--4~keV and 8--15~keV
and emission in the 5--7~keV band containing the line
in a long {\it RXTE} observation of MCG-6-30-15.
He did not find any reverberation delays longer than 500 seconds.
This result, along with X-ray variability data allowed him to
conclude that the mass of the black hole is $\sim
10^6-10^7\,M_{\odot}$ in agreement with Novak~\& Chiang~(1999)
result. A multiwavelength study of MCG-6-30-15 (Reynolds et al.,
1997) shows that the bolometric luminosity of the object is $\approx
8\cdot 10^{43}\,\mbox{erg s}^{-1}$. Thus, the mass of
$10^6\,M_{\odot}$ implies that the accretion luminosity is roughly 60\% of the
Eddington luminosity. This estimate provides some grounds for the
hypothesis that the extended edges of the line are caused by
turbulence in an accretion disk.

If one believes that the luminosity of the accretion disk is  60\% of 
the Eddington limit, one can conclude that the black hole in 
MCG-6-30-15 cannot rotate at the maximum rate, since disks around 
fast rotating Kerr black holes do not have turbulent velocities high
enough to make a $30^\deg$ inclination angle consistent with $g_{\rm max}$ for 
the ``Int'' data point. Together with the lower bound for $a_{\ast}$ from
the $g_{\rm min}$ values of the reddest observed line profiles (see 
below) this leads to the estimate of  $a_{\ast}\approx 0.3$. 
It is interesting to point out that knowledge
of the accretion disk luminosity can lead to a value of $a_{\ast}$
from the position of the blue edge of the line. 
In the future, with improved understanding 
of the inner disk structure, one can hope to use
the shape and position of the blue edge of the profile to obtain 
a better quantitative estimate of $a_{\ast}$.

Still, we cannot rule out that highly ionized regions of the disk,
emitting iron lines at $6.7\,\mbox{keV}$ and $6.97\,\mbox{keV}$,
can also produce blue wing of the line.  As seen in Figures~\ref{fig2}
and~\ref{fig6}, when the continuum was brighter (see the 97a data
point from Iwasawa et al.~1999), the $g_{\rm max}$ was smaller than when the continuum was
dimmer (97b data point). The data
sets can be explained by asserting that in the 97a
case the bright flare occurred close to the innermost region of the
disk, and more distant regions of the disk were not sufficiently well
illuminated to produce an extended blue wing. 
If highly ionized iron were skewing our inference of the blue
edge, it would presumably do so in a manner that would be more noticeable
during the bright flare. 
It does not seem that such a contribution spontaneously appears during the 
bright flare, since this would push the true $g_{\rm max}$ lower than
we have measured, and the 97a data would then imply a lowered inclination
angle bound which is inconsistent with the other datasets.
In the absence of higher quality data, we can only say
that contamination by a 6.7~keV line occurs in both cases or not at all.

The position of the Iwasawa et al. (1997) DM and the Iwasawa et
al. (1999) 97a point in the $g_{\rm min}$--$g_{\rm max}$ diagram
indicate that at the level of $2\sigma$ confidence there is emission
coming from below $6 R_g$, the innermost stable orbit of a
Schwarzschild black hole. Figures~\ref{fig2} and~\ref{fig6} show that
this conclusion is true regardless of the value of the black hole spin
and thickness of the accretion disk.  This can be considered as
robust, model-independent evidence that the central object in
MCG-6-30-15 is emitting from regions in the strong gravitational field
of a black hole.  The most probable innermost radius of
emission as determined by the 97a and DM points with an assumed inclination
angle of $30^\deg$ is 4--5$R_g$, and it is still within 6$R_g$ for our higher
inclination angle estimates. 
The location of the inner
edge of the disk at these radii corresponds to the rotational
parameter $a_\ast$ to be in the range between $0.29$ and $0.56$. Since
there is no significant iron absorption edge in the DM and 97a profiles,
it is unlikely that any free-falling gas below innermost stable orbit
can account for the observed extended red tails (Young,
Ross, \& Fabian, 1998).  
If the red tails of the DM and 97a profiles are
produced by the emission of ionized iron, this will decrease $g_{\rm
min}$ by a few per cent and make the estimates of the innermost radius
of the disk even smaller and the rotational parameter $a_\ast$ even
higher. The caveat here is that we are able to determine only an upper bound
for $g_{\rm min}$ of DM and the best nonlinear fit for 97a red edge
falls close to the lower end of the error interval (Fig.~\ref{fig6},
section~2). We could be missing extended red tails
in the profiles as a result of
large scatter and observational errors. 
Thus, it is reasonable to view our results for $a_\ast$ as only
a lower bound, i.e., that the black hole in MCG-6-30-15 rotates faster
than about $a/M=0.3$.

\subsection{NGC~4151}

NGC~4151 is a bright nearby ($z=0.0033$) Seyfert~1.5 galaxy. It
exhibits a broad K$\alpha$ line profile very similar to the one
observed in MCG-6-30-15 but with better signal-to-noise (Wang et
al., 1999).  The observed biconical geometry of the narrow line
[O~\uppercase{iii}]~$\lambda$5007 region (Evans et al., 1993) suggests
edge-on geometry of the accretion disk. The best estimate of the angle
between axis of the emission line cone and line of sight is $65^\deg$. 
We may be tempted to adopt this value of inclination for the
X-ray emitting disk as well, however, better constraints
come from the geometry of the observed radio jet (Pedlar et al.~1993)
which presumably originates in the inner region of the AGN.
Unlike Evans et al.~(1993), Pedlar et al. (1993) consider a geometry of the
narrow line region in which the [O~\uppercase{iii}] emission comes from a
stripe where the ionization cone touches the galactic disk. This
hypothesis is justified by a comparison of the velocities of
the [O~\uppercase{iii}] region and those of neutral hydrogen, which
require the [O~\uppercase{iii}] region to participate in the galaxy
rotation and, thus be close to the plane of the galaxy. If the axis
of the radio jet coincides with the axis of the ionization cone, then
one can infer that the angle between the radio jet
and the line of sight to be $40^\deg$ (which turns out to be consistent with
the observed relativistic beaming of $v=0.15 c$). Naturally, one expects the
radio jet to be perpendicular to the plane of the inner part of the
accretion disk where the collimation process occurs. This would then
give an estimate of $40^\deg$ for the inclination angle of X-ray
emitting part of the accretion disk.

Our $g_{\rm min}$--$g_{\rm max}$ findings for NGC~4151 are plotted in
Figures~\ref{fig1} and \ref{fig5}. 
The $2\sigma$ lower bound for the inclination angle of the disk
derived from the thin disk model and a nonlinear edge fit (Fig.~\ref{fig5})
is approximately $28^\deg$. The best-fit frequency extrema point in
Fig.~\ref{fig5} gives a most probable inclination of $35^\deg$. The upper
$2\sigma$ limit is $47^\deg$. All of these numbers are almost independent of
value of $a_{\ast}$. Polynomial fitting (Fig.~\ref{fig1}) gives
somewhat larger values for $g_{\rm max}$; the corresponding results for
the inclination angle are: the lower $2\sigma$ bound is $40^\deg$, best fit
value is $48^\deg$, and the upper limit is $60^\deg$. The bounds differ by
no more than $5^\deg$ for different values of $a_{\ast}$. The line
emission comes from a disk radius beyond $6 R_g$, with the most
probable location of the innermost emission region at
8--10$R_g$.  The turbulent disk model gives lower inclination angles
and slightly (by approximately $1 R_g$) higher radii of the location
of emissive spots.

Wandel, Peterson,~\& Malkan~(1999) determine virial masses,
emission-line region sizes, and the flux of ionizing continuum of AGNs
using reverberation and photoionization techniques.  For NGC~4151 they
find the mass of the black hole to be $1.2$--$2.2\,\cdot 10^7\,M_{\odot}$,
the ionizing luminosity $5\cdot 10^{42}\,\mbox{erg s}^{-1}$, which
corresponds to the ratio of ionizing luminosity to the Eddington
luminosity as $3\cdot 10^{-3}$. 
The ionizing luminosity is presumably comparable to the bolometric
luminosity, since a large part of the AGN energy is expected to be
radiated in the UV band. Thus, it is unlikely that the thickness of
the disk has any effect on the iron line profile in NGC~4151.

Our analysis excludes the value of $65^\deg$ for the inclination angle of
the disk and is in good agreement with the value of $40^\deg$ deduced by
Pedlar et al.~(1993).  The unshifted main core of the line can be
attributed to some relatively narrow line component whose origin is
not the accretion disk. This component can be due to reflection of 
line photons from a surrounding torus and cool corona above the accretion
disk (Poutanen et al. 1996).

\subsection{NGC~3516}

NGC~3516 is another close Seyfert~1 galaxy ($z=0.009$) which is known
to have a broad, skewed K$\alpha$ iron line (Kriss et al., 1996;
Nandra et al., 1997b; Nandra et al., 1997c; Nandra et al., 1999). We
used the most recent observations by Nandra et al.~(1999) for the
application of our method.  The $g_{\rm min}$-$g_{\rm max}$ point for
the line profile for the whole observation (Fig.~1 in Nandra et al.,
1999) as well as points for the line profile at each time interval
(Fig.~3 in Nandra et al., 1999) were calculated. Figure~\ref{fig3} shows
the results of polynomial fitting of lines edges, while
Figure~\ref{fig7} represents the results of nonlinear fitting. One can
see that the positions of the $g_{\rm min}$-$g_{\rm max}$ points obtained
by these two different methods fall within each other's error boxes,
though the errors of the nonlinear model are larger than polynomial
fitting. To be conservative, we use the nonlinear fits
(Fig.~\ref{fig7}) in our analysis. From the position of points P2, P4,
and P5 we conclude that the lower limit on the inclination angle of
a thin Keplerian disk is $27\pm 4^\deg$. The best-fit locations in
Fig.~\ref{fig7} indicate an inclination angle of the disk $43\pm 4^\deg$.
These estimates are smaller for a rapidly rotating black hole but only by a
few degrees.

We cannot put an upper limit on the inclination angle of the disk in
this case.  Nandra et al.~(1999) fit an 
integrated state (``Int'')
profile and obtain the following values for the inclination of a thin disk:
In the Schwarzschild model, $i={35^{+1}_{-2}}^\deg$, and for the
extreme Kerr model
$i={0^{+19}}^\deg$. We see that the appearance of blue wings in profiles
 P2, P7,
and P4 causes our estimates of the inclination angle to increase
compared to the fits from the integrated line profile.  This discrepancy
might be understood from Figure~1 of Nandra et al.~(1999), showing
that their fits for Schwarzschild and Kerr geometries have red wings
which extend beyond the actual red edge of the line
profile. Our determinations of the red edge of the Int profile give:
linear fitting (Table~1) $3.53\pm 0.19\,\mbox{keV}$; nonlinear best
fit (Table~2) $3.34\,\mbox{keV}$; and nonlinear fits to Monte-Carlo
scattered points (Table~2) $3.98\pm 0.45$. For the nonlinear fits we
used only data points from channels 5 to 24, excluding the depression below
$3.5\,\mbox{keV}$ which is caused by absorption. 
For comparison,
the fitting domain of Nandra et al.~(1999) extends below $3\,\mbox{keV}$.

The inclination angle $i={35^{+1}_{-2}}^\deg$ obtained by Nandra et
al.~(1999) for the Schwarzschild model is still consistent with our lowest
limit, however, their estimate for the Kerr model is
lower than any of our limits.  Thus, our line edge
determinations for time-resolved observations (P1-P8) favors
a Schwarzschild black hole --- the best fit to a Kerr model by Nandra et
al.~(1999) simply cannot account for blue wings of the line in P2, P4, and
P5 time intervals.  
These transient blue wings may be caused, for
example, by a temporary enhancement of irradiation of outer parts of the
disk during the P2, P4, and P5 observation
intervals. In the combined set of data, this enhancement of the blue wing
gets averaged out and contributes to the change (decrease) of the
index of the best-fit power law model of the continuum. 

NGC~3516 observations do not show a red tail as extended as in
MCG-6-30-15. Large uncertainties in the data points in
Fig.~\ref{fig7} prevent us
from placing tight constraints on the radial location of the emitting
region or from deriving any conclusions about the possible extent of
the accretion disk below $6 R_g$.  The point with the lowest upper bound
on $g_{\rm min}$ is P2. Using the square of errors for that point one
can only conclude that the disk in NGC~3516 must be extended below $12
R_g$ at a $2\sigma$ level of confidence, if one adopts $i=35^\deg$ for the
disk inclination.

Another possibility is that a thick disk model (Figs.~\ref{fig7}b
and~\ref{fig7}d) could naturally explain the blue extended
wings of the line without changing the fit to the core. 
Analysis of unevenly sampled X-ray luminosity data on NGC~3516 led
Edelson~\& Nandra~(1999) to obtain a rough estimate of the mass of the
central black hole in NCG~3516 of $10^7\,M_{\odot}$. Infrared, optical
and near UV fluxes from the NGC~3516 nucleus total $\approx 3\cdot
10^{-10}\, \mbox{erg s}^{-1}\mbox{cm}^{-2}$ (NASA/IPAC Extragalactic
Database), while the far UV flux is $\approx 6\cdot
10^{-11}\,\mbox{erg s}^{-1}\mbox{cm}^{-2}$ (Goad et al.~1999), and the
soft X-ray flux (NASA/IPAC Extragalactic Database) is $3\cdot
10^{-12}\, \mbox{erg s}^{-1}\mbox{cm}^{-2}$. This sums up to give a
lower bound for the bolometric luminosity of $\approx
10^{44}\,\mbox{erg s}^{-1}$ or the ratio $L/L_{\rm ed}\approx
0.08$. 

If the above luminosity estimates are accurate, then the effects of turbulent
broadening on the line profile are negligible. However, $L/L_{\rm ed}$
can be a few times larger, with the main uncertainty coming
from poor knowledge of the mass of the central black hole. When
results of reverberation studies for NGC~3516 become available,
we hope to obtain a better estimate of the central black hole mass and
to determine $L/L_{\rm ed}$ more accurately. 
Still, Figure~\ref{fig7}d indicates
that even a ``maximally'' thick disk with $L=L_{\rm ed}$ cannot make the
Nandra et al.~(1999) Kerr best fit value of $i={0^{+19}}^\deg$ to be
consistent with $g_{\rm max}$ measurements for P2, P4, and P5
intervals (all the points of the curve $i=19^\deg$ are below the error
squares for P2, P4, and P5). In the case of a Schwarzschild model,
comparison of Figs.~\ref{fig7}a and~\ref{fig7}b shows that the ratio
of $L/L_{\rm ed} \approx 0.3$ is enough to raise the curve $i=35^\deg$
such that it will be consistent with the best fit points for all
observational intervals except P5 and P8. The inability of the Kerr
thick disk model to account for the blue wings of the line strengthens
our conclusion that the black hole in NGC~3516 should be rotating
slowly

\section{Summary}

In this work we suggest a method which allows us to put bounds on the
geometry and kinematics of the line emitting surface of a relativistic
accretion disk around a supermassive black hole.  Our technique is not
sensitive to models of the illuminating source, the distribution of
illuminating radiation over the disk surface, or the angular
distribution of the reflected line in the rest frame of the reflecting
material.  We use our method to determine the positions of the red and
blue edges of observed emission lines and extract information from
these positions by comparison with the results of a fully relativistic
ray-tracing code.  The code generates maps of line-frequency extrema,
and it can be used to demonstrate explicitly the differences which can
arise as a result of black hole spin and turbulence in an accretion
disk.
 
Conventionally, the only way to identify black hole rotation
is to find emission from within the innermost stable orbit
of a nonrotating black hole. Our method illustrates that if one has
information about the geometry of the accretion flow from some
other observation such as the inclination angle of a jet, along
with an estimate of the luminosity relative to Eddington, then
a new measure of the hole's spin may be available.
We can thus break the ``degeneracy'' that exists
between rotating and nonrotating black holes in a new way.

Certainly, our method cannot be a substitute
for the fitting of the whole shape of the line assuming a particular 
emissivity law (power law axisymmetric in most works) and it cannot provide
interesting information about the emissivity law itself. However, it can
provide bounds on the geometry of the disk and the angular momentum of the
black hole, which any emissivity model should satisfy. Therefore, it is 
important to emphasize the position of the line edges and to test model
parameters obtained as a result of line profile fitting against the 
constraints provided by our method. 

A drawback of the method are uncertainties in the determination of
the position of the edges. A main focus of this paper is
to find a reasonable edge detection algorithm to replace the
conservative edge-boundary limits found by Bromley, Miller \& Pariev (1998).
We perform Monte-Carlo 
simulations of the data set using fits to the line edges
with both a nonlinear sharp-edge model and a polynomial model which
makes no a priori assumption about the shape of the line profile. We find
reasonable agreement between models, both in terms
of best-fit values and the error distributions.

Another drawback of the 
method is possible contamination with iron K$\beta$ line, nickel K$\alpha$
line, $6.7\,\mbox{keV}$ and $6.97\,\mbox{keV}$ K$\alpha$ lines of highly 
ionized iron. We made corrections to profiles for the iron K$\beta$ 
line using a K$\beta$-to-K$\alpha$ yield ratio
and for the nickel K$\alpha$ line using a fiducial yield of $0.06$ relative to
iron K$\alpha$ (see George \& Fabian~1991). For 
quality of spectra currently available, these corrections are small.

We illustrate frequency extrema method by applying it to the 
Seyfert
galaxies MCG-6-30-15, NGC~4151, and NGC~3516. The results for MCG-6-30-15
show that the commonly assumed 
inclination angle of $30^\deg$ for a thin accretion disk is inconsistent 
with the position of the blue edge of the line at a $3\sigma$ level. 
The thick turbulent disk model can remedy this discrepancy, since it leads 
to appearance of a smooth blue wing in the line while the changes to 
the main profile are not so large (Pariev~\& Bromley, 1998). Recent
estimates of the mass of the black hole in  MCG-6-30-15 favor a 
luminosity ratio $L/L_{\rm ed} \sim 0.6$, which is enough to account 
for the $g_{\rm max}$ for all observations of the iron line in MCG-6-30-15
published to date. Furthermore,
frequency extrema lead us to conclude that the black
hole in MCG-6-30-15 must be rotating with at least $a/M=0.26$.

For NGC~4151 our method excludes face~on and edge~on geometries, while
giving bounds for the inclination angle of X-ray emitting inner disk
of $50\pm 10^\deg$. These bounds are consistent with the models of  
the ionization cone grazing the disk by Pedlar et al.~(1993) 
and two X-ray emitting disks of 
Wang et al.~(1999). However, our bounds are not consistent with 
the $i=65^\deg$ geometry
assumed by Evans et al.~(1993) based on the observation of the
biconical geometry of the narrow line
[O~\uppercase{iii}]~$\lambda$5007 sources. Because of very low ratio of
$L/L_{\rm ed}$, effects of turbulence in this source are negligible.

We find that frequency extrema results combined with line profile fitting
by Nandra et al.,~(1999) favors
Schwarzschild vs. Kerr model for 
NGC~3516. Although $L/L_{\rm ed} \sim 0.08$ for this source, the determination
of the mass of the black hole is very uncertain. The possibility exists 
that the thick disk model may be relevant for NGC~3516 and could 
explain the blue wings of the line, observed during monitoring of 
this source by Nandra et al.~(1999). 

Further reverberation mapping results will provide better estimates of
the mass of the black holes in nearby Seyferts and allow the thick
turbulent disk model to have more predictive power for each individual
source. Certainly, going beyond our simplifying assumptions and
incorporating results of modeling of X-ray reflection spectra from 
the disk surface
would be the next step to interpret higher quality spectral data from
XMM and Constellation-X missions.

\acknowledgments

We are grateful to K.~Nandra and J.-X.~Wang for supplying us with the
digital versions of iron line profiles.
Many thanks go to S.A.~Colgate for constant support and 
encouragement during work on this project. The comments of the 
referee improved presentation of this work.
The SGI/Cray supercomputer used in this research was provided through
funding from the NASA Offices of Space Sciences, Aeronautics, and
Mission to Planet Earth.  This research made use of the NASA/IPAC
Extragalactic Database (NED), which is operated by the Jet Propulsion
Laboratory, California Institute of Technology, under contract with
NASA.  
Partial support from the NASA Astrophysics Theory Program and 
U.S. Department of Energy through the LDRD program at Los Alamos National
Laboratory is also acknowledged.

\newpage

\figcaption[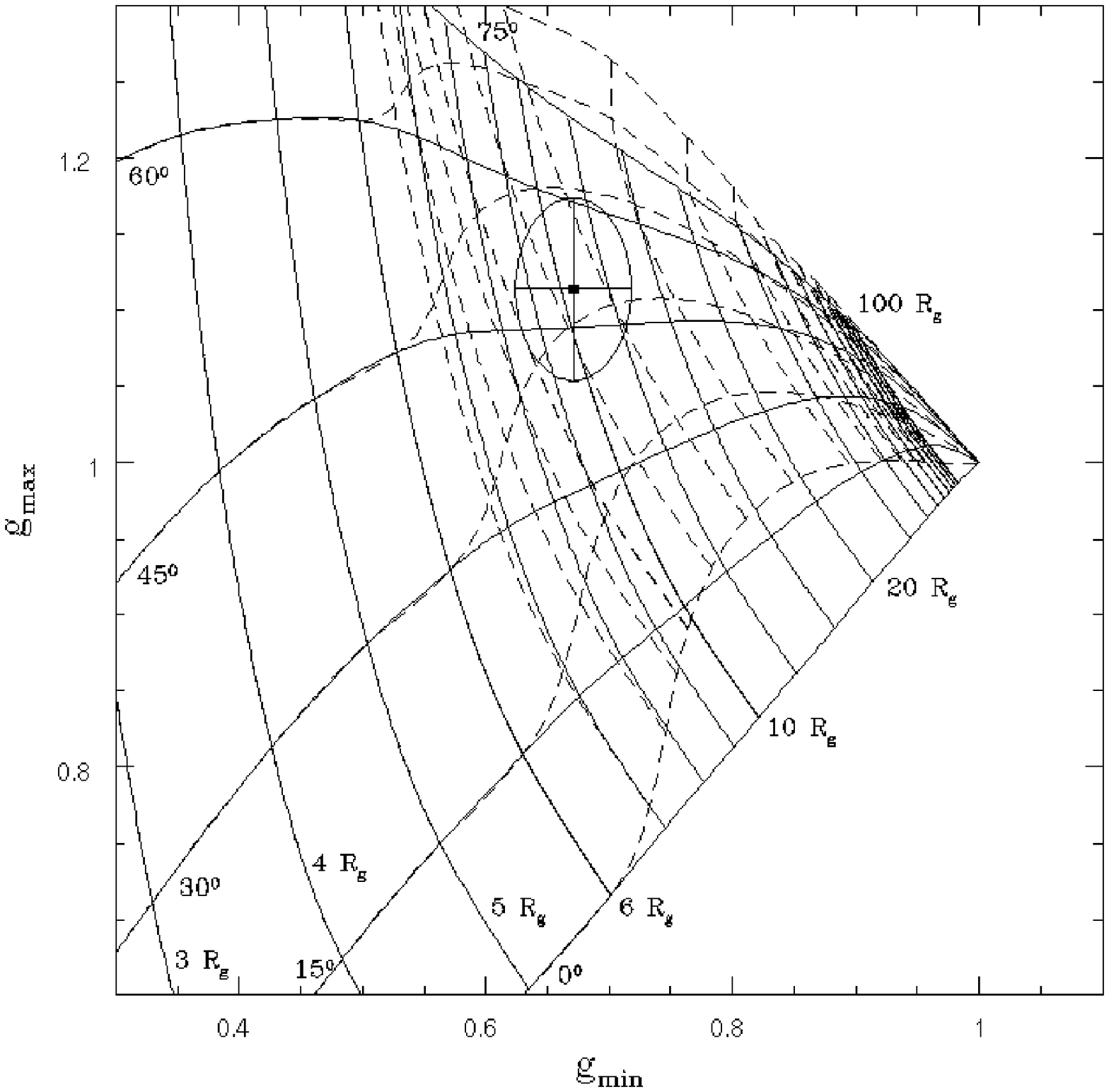, 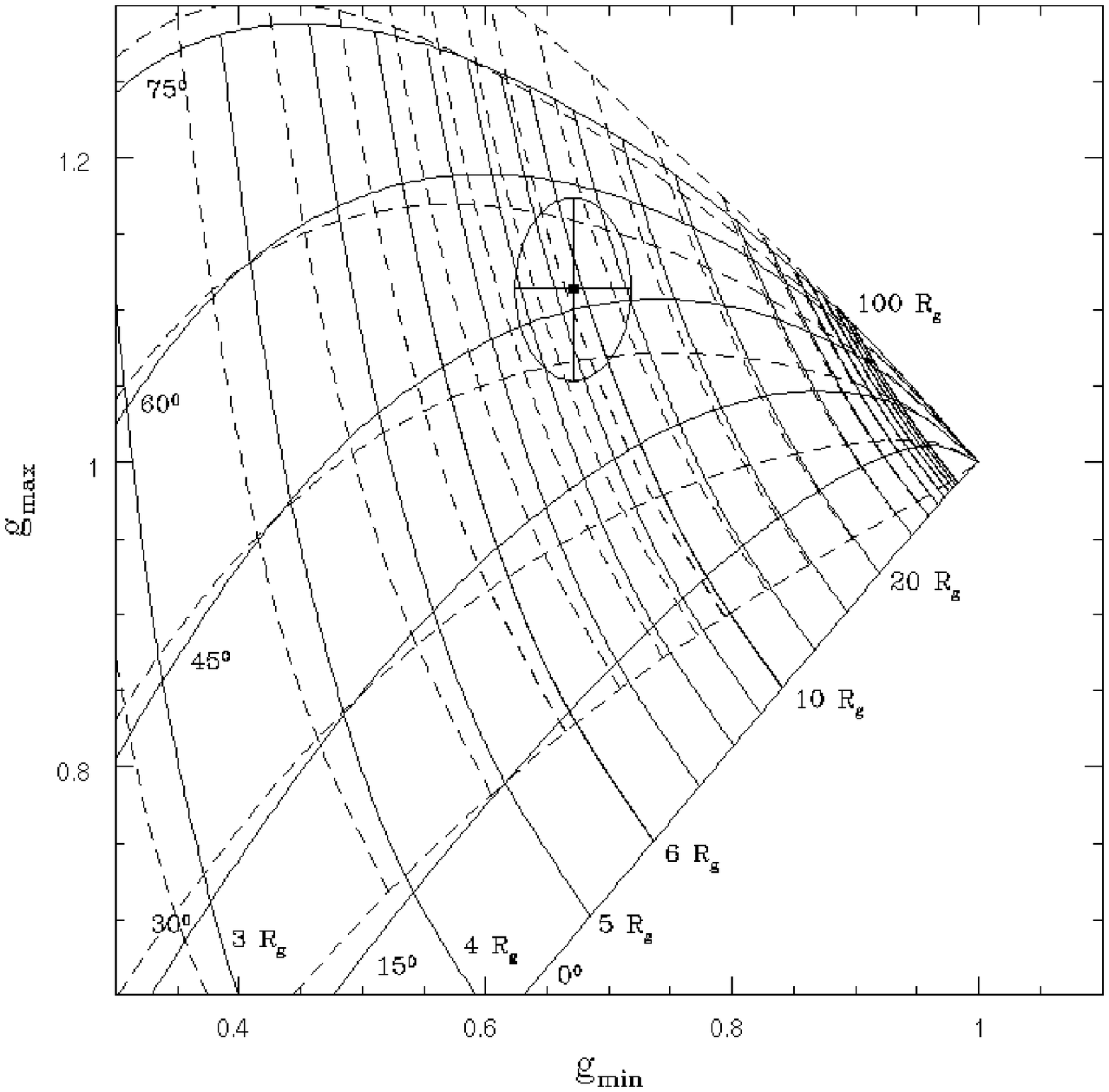]
{Maps of minimum and maximum redshifts for the case of Schwarzschild
(a) and extreme Kerr (b) black holes. The grid of constant inclination angle 
(approximately horizontal lines) and equal radii (lines in vertical direction)
are plotted. Solid lines correspond to the thin Keplerian
disk model; dashed lines are for the turbulent thick disk model with 
Eddington luminosity
$L=L_{edd}$. In the region with radii $r<6 R_g$ in the Schwarzschild case 
the gas was 
considered to be free-falling.
The point is for NGC~4151 data by Wang et al. (1999).
Error ellipse of polynomial fits 
is drawn at $2\sigma$ level contour. The scale on plots (a) 
and (b) is the same as well as on all $g_{min}$--$g_{max}$ plots in this
work. \label{fig1}}

\figcaption[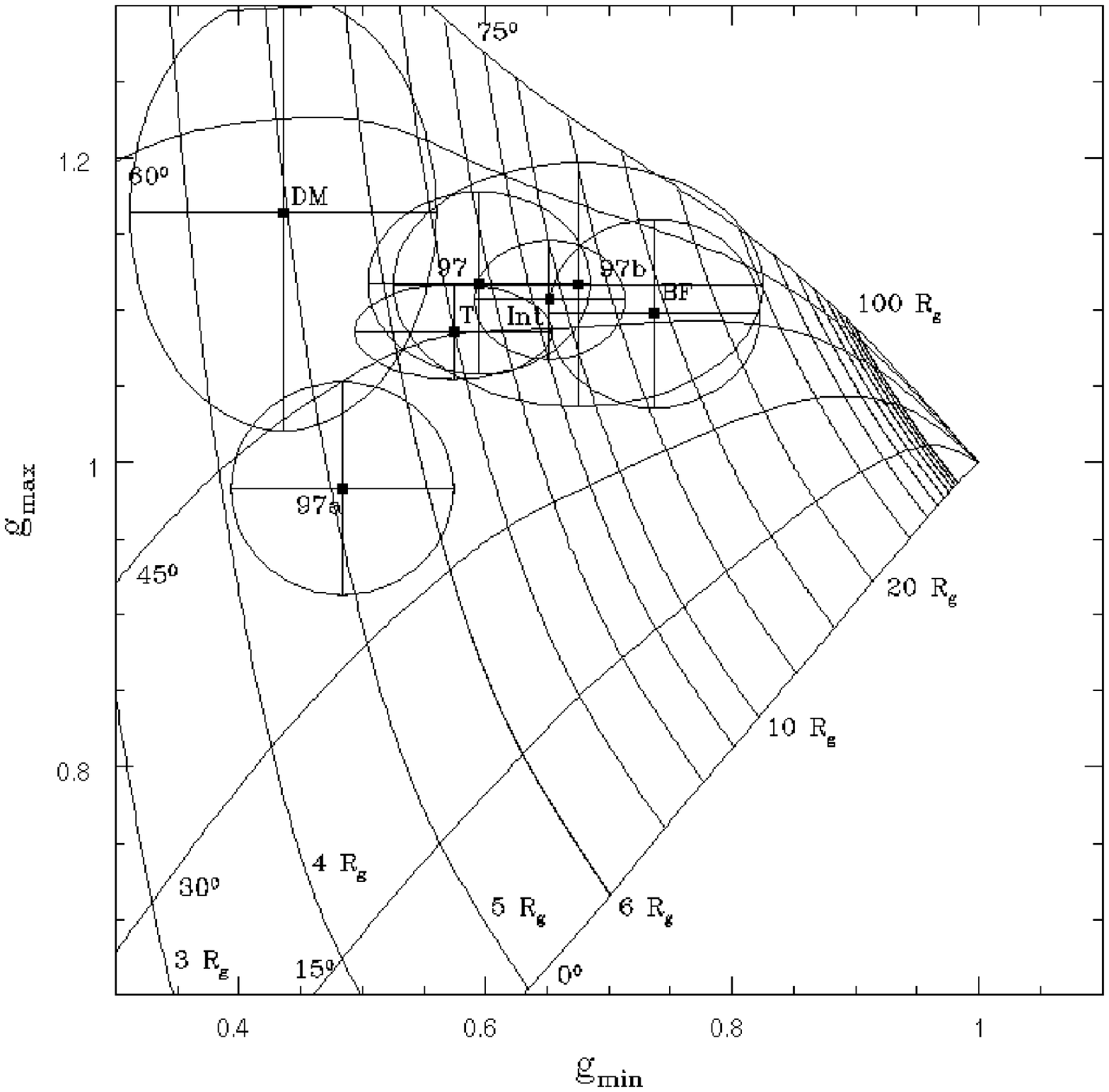, 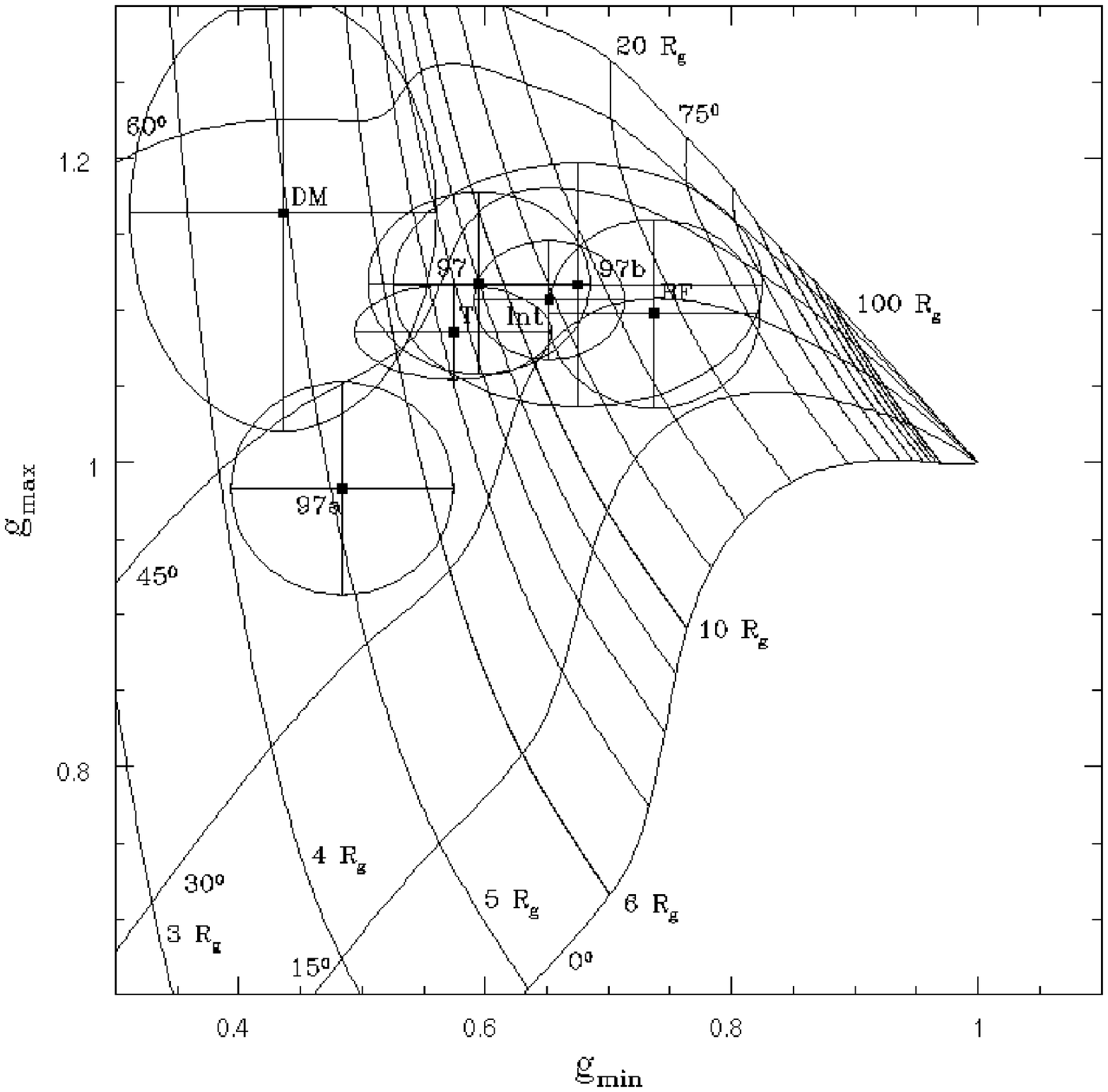, 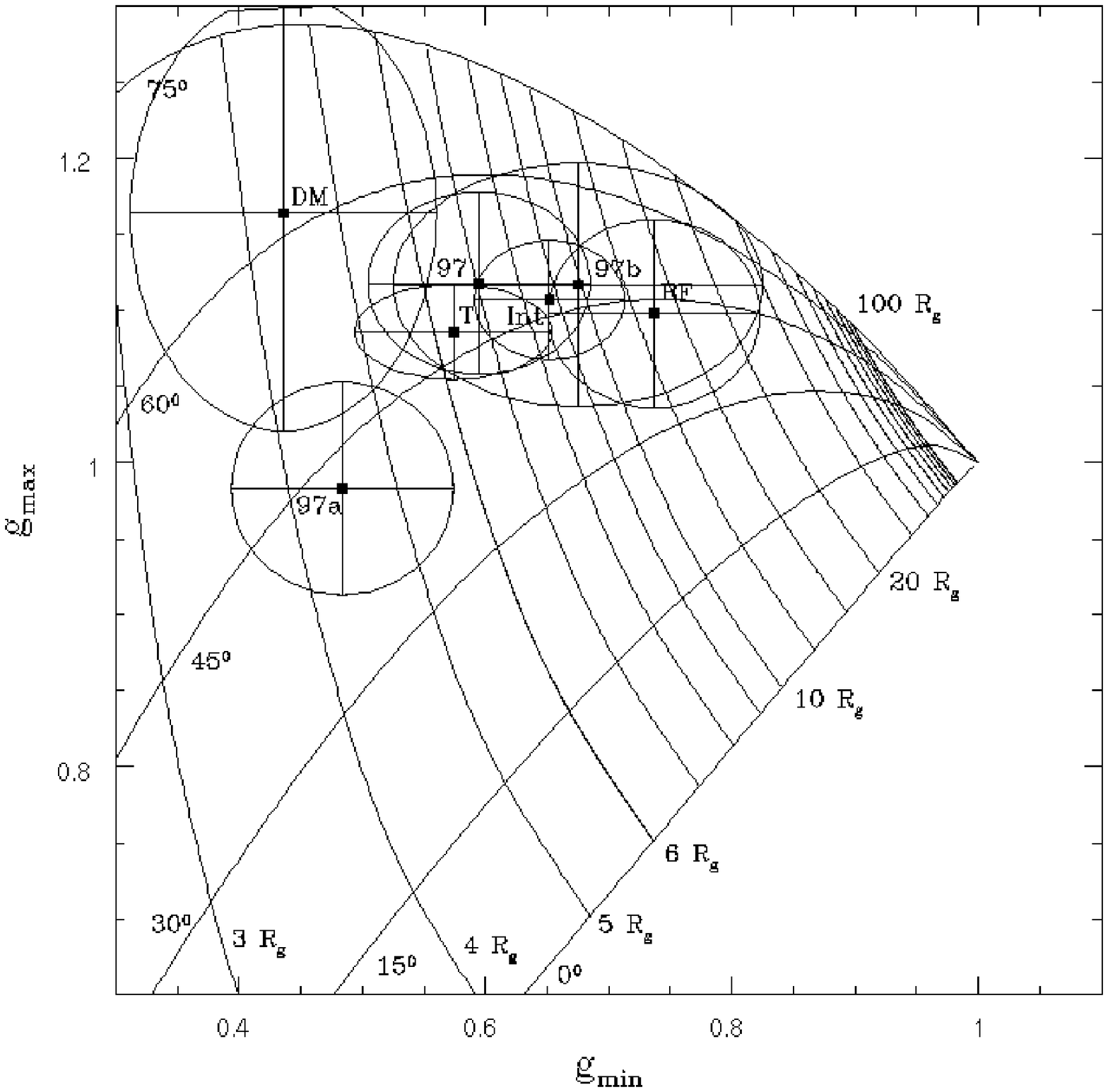, 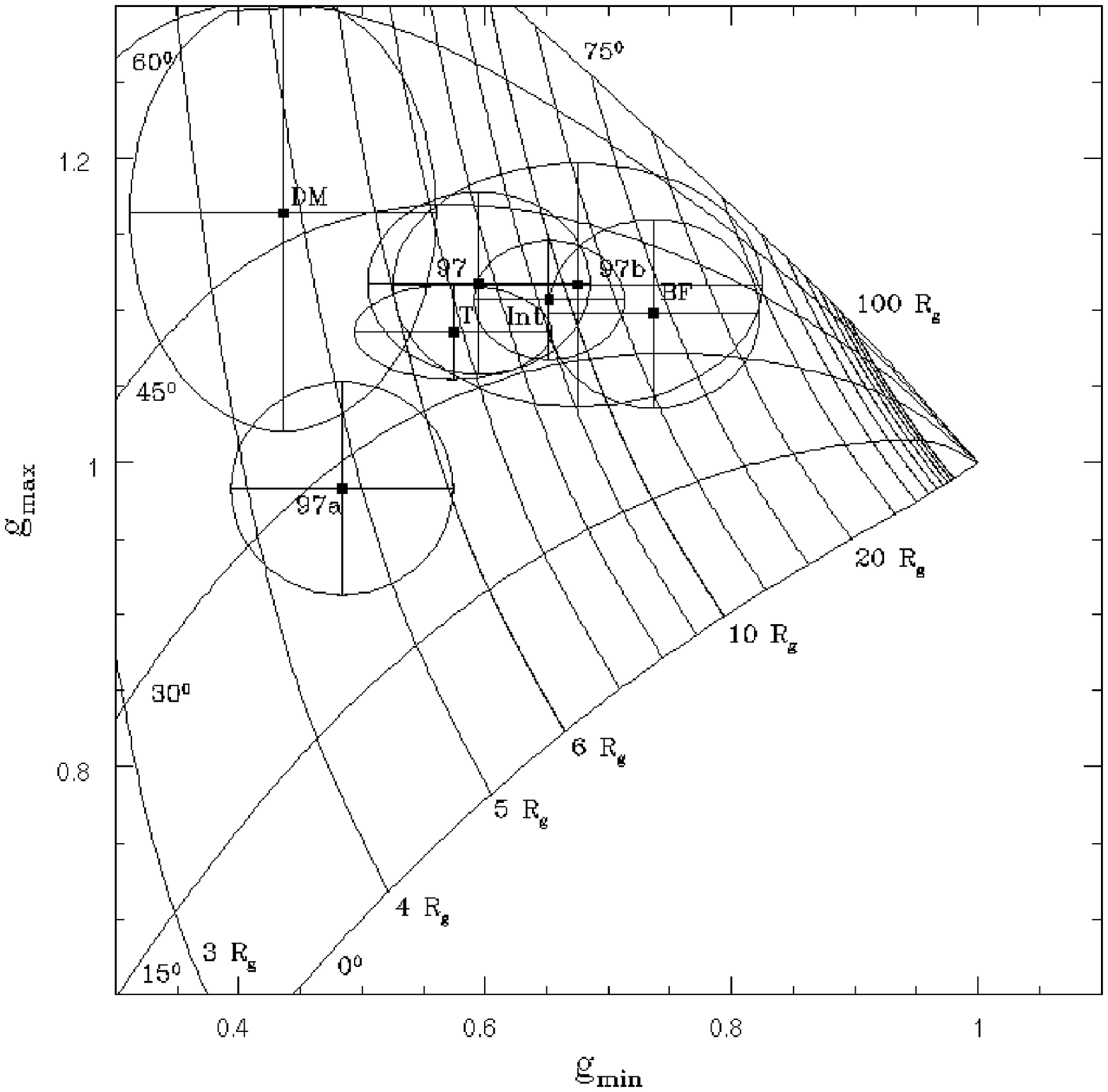]
{Data from MCG-6-30-15 plotted on a maximum and minimum frequency 
shift diagram. Error ellipses of polynomial fits 
are shown at a $2\sigma$ level contour.
Observational data points on all four panels are the same. 
The point marked as T stands for the Tanaka et al. (1995) data.
Int, BF and DM denote 
intermediate, bright flare and dark minimum spectra from 
Iwasawa et al. (1996). Points 97, 97a and 97b are 
for the average profile (97), bright flare subset (97a) and minimum subset
(97b) from Iwasawa et al. (1999).  The panels differ by the model used 
for the disk--black hole system. The top two panels are for a Schwarzschild 
black hole: Fig.~(a) is for thin disk model, and  Fig.~(b) is for
thick disk model. The bottom two panels are for extreme Kerr cases: Fig.~(c)
is for the thin disk model, Fig.~(d) is for the thick disk model.
\label{fig2}}

\figcaption[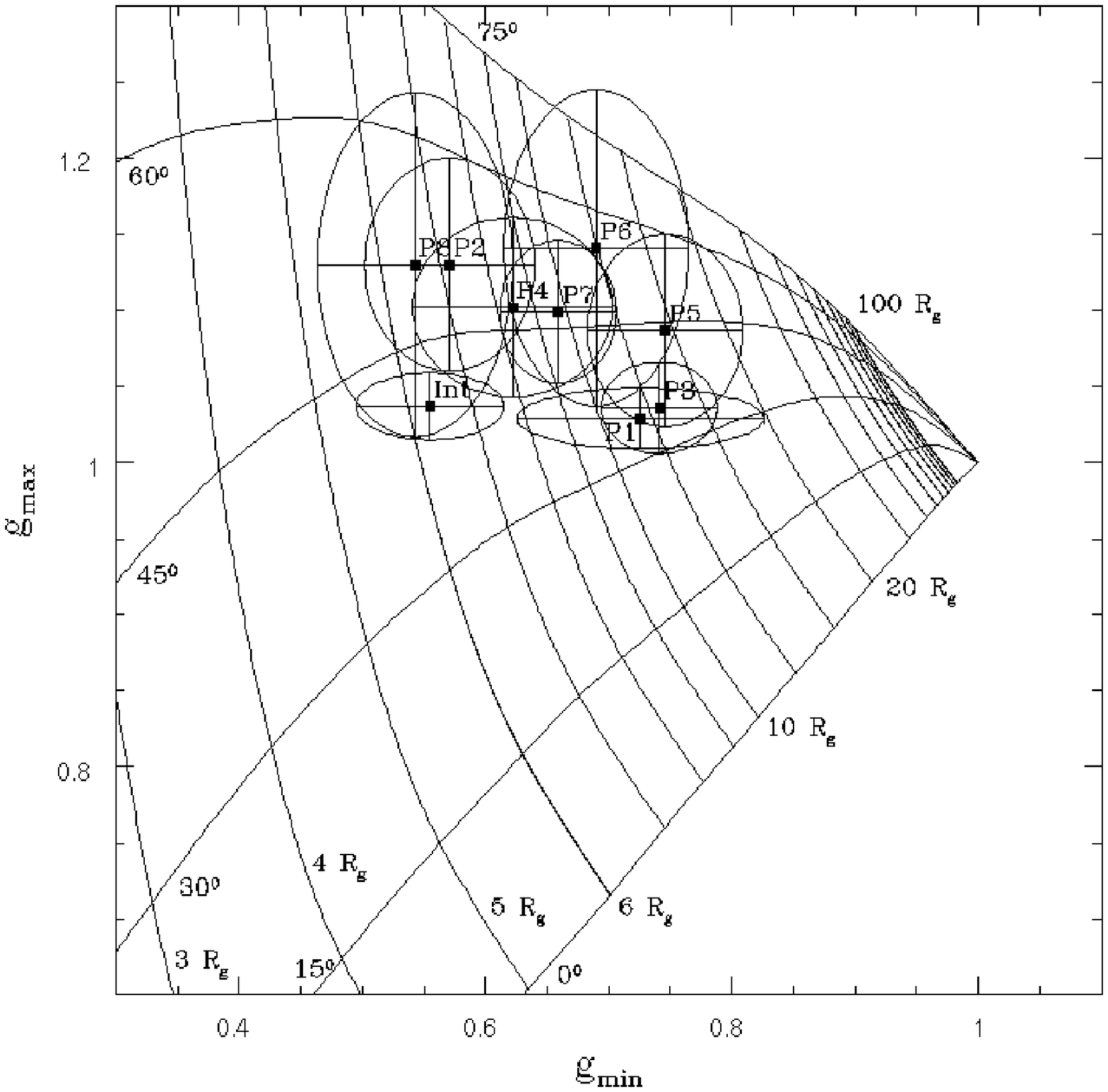, 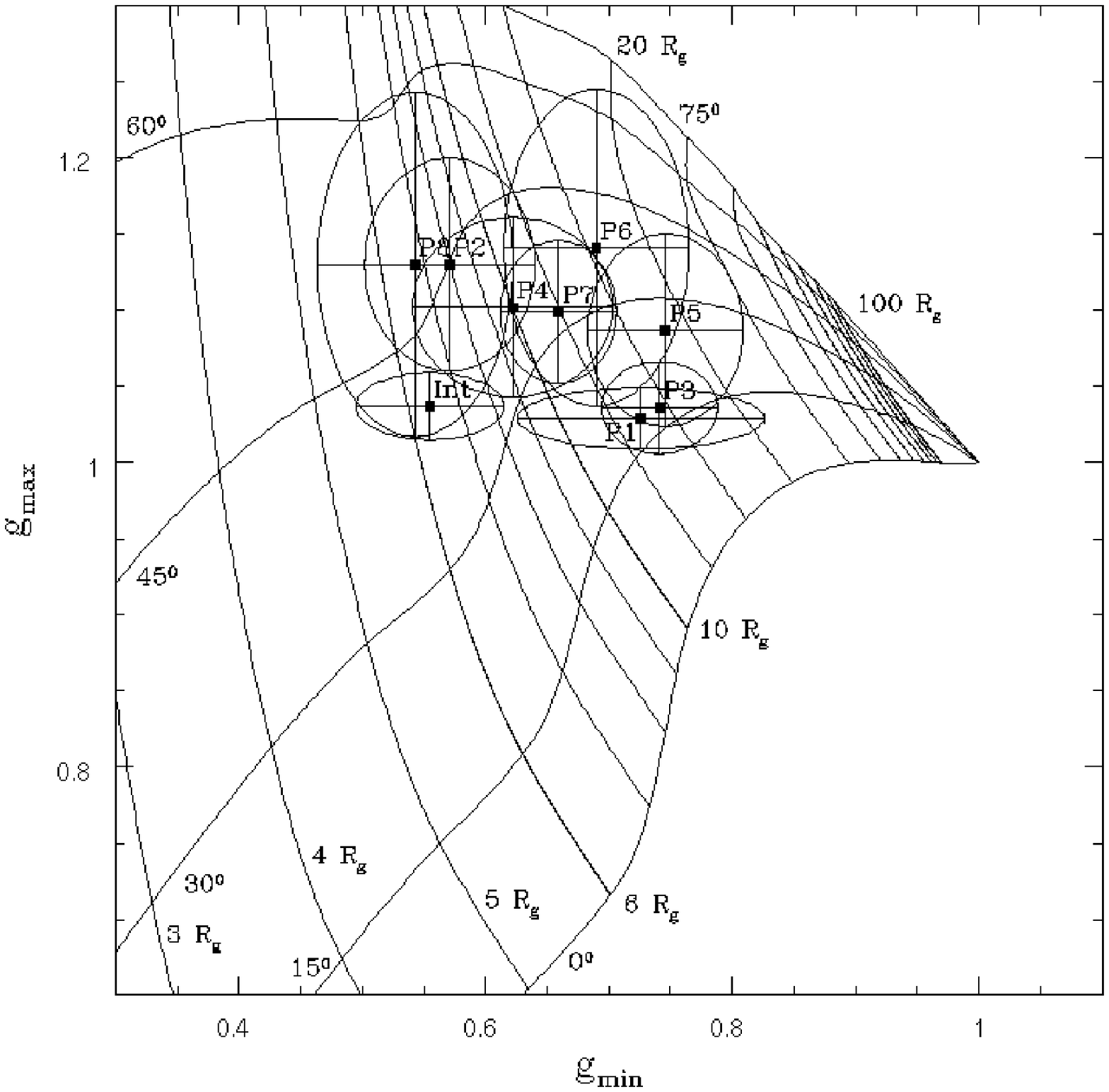, 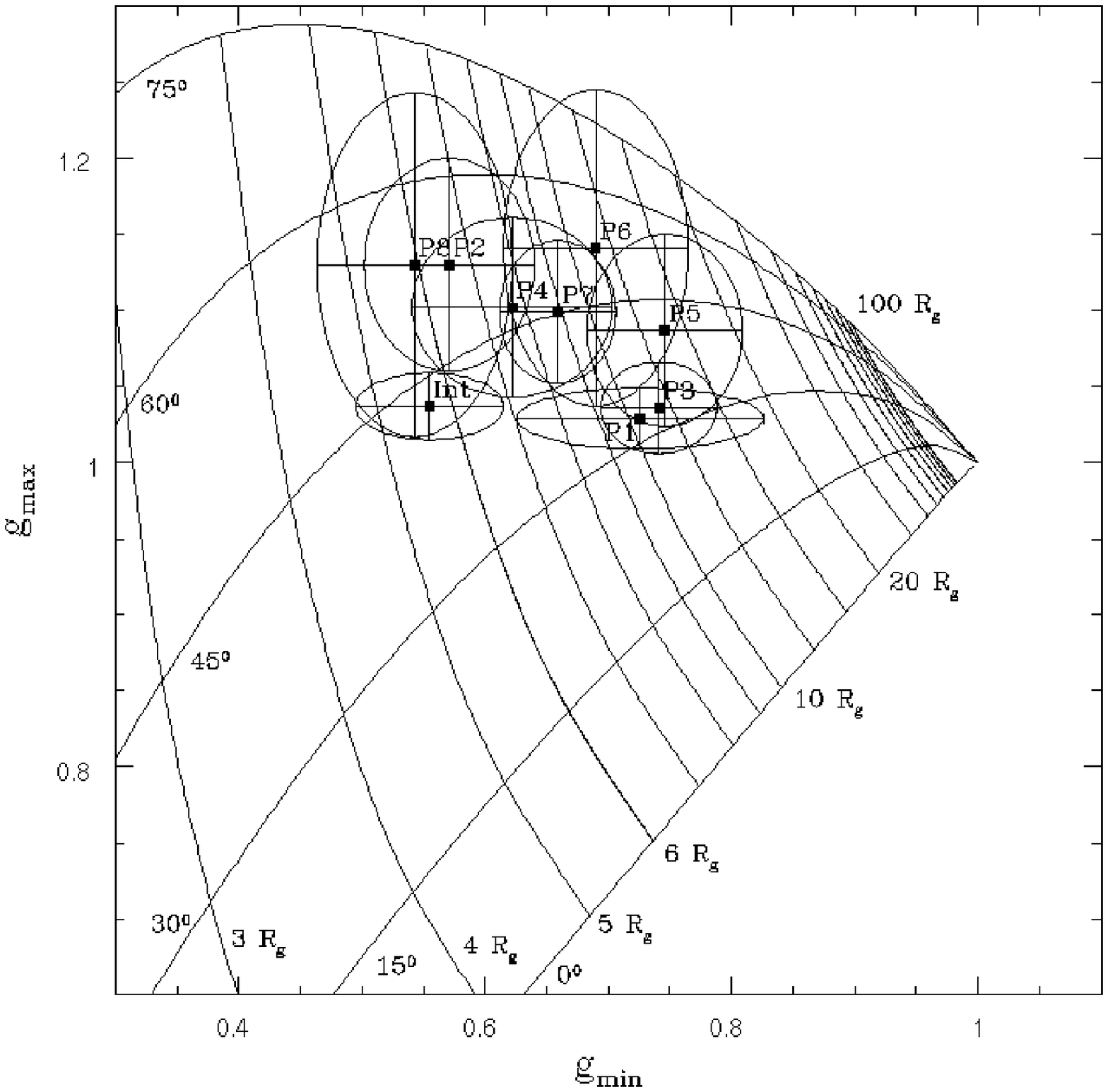, 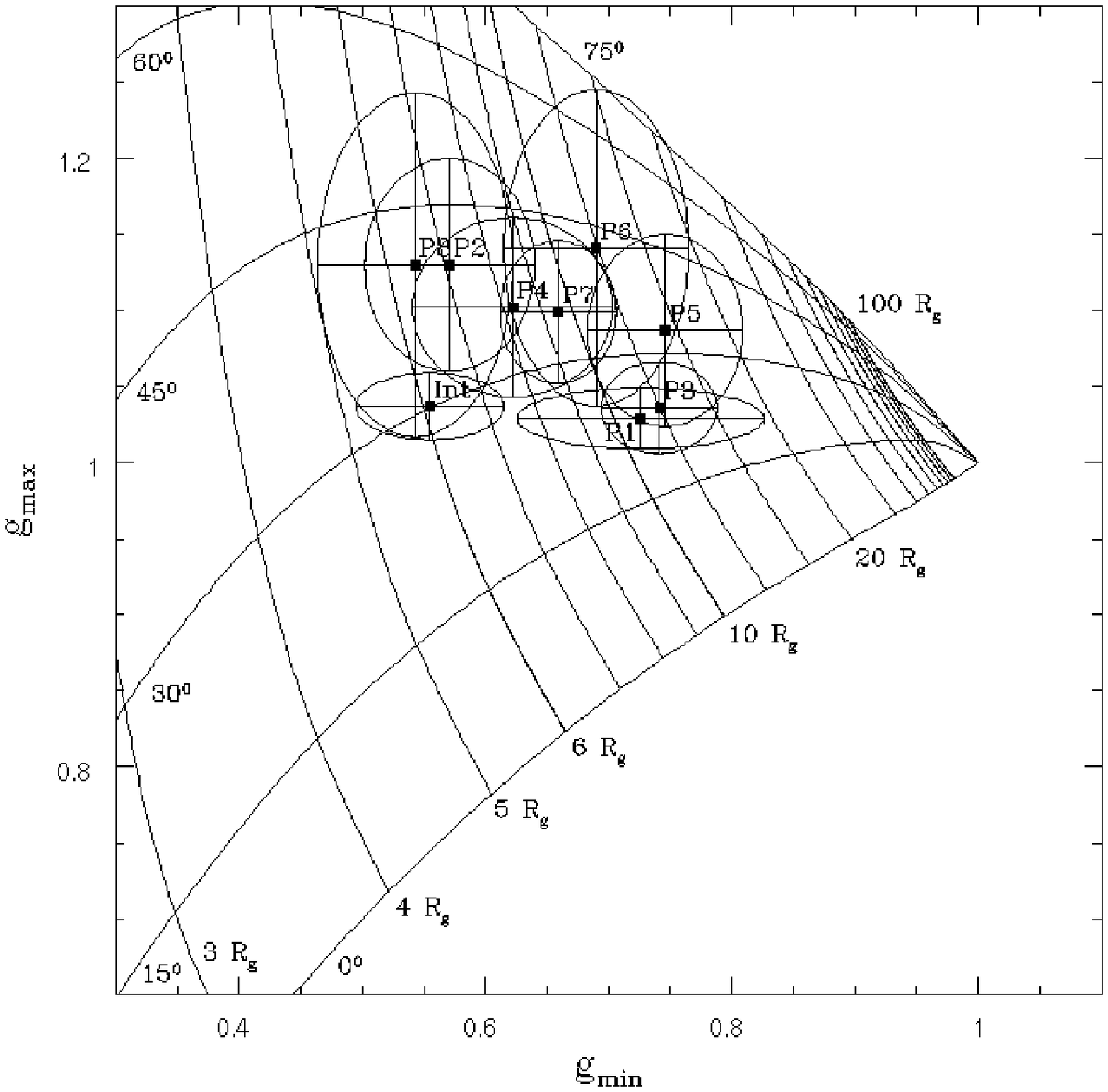]
{Data from NGC~3516 plotted on maximum and minimum frequency
shift diagrams. Error ellipses of polynomial fits 
are shown at a $2\sigma$ level contour.
Observational data are the same on all panels and are taken from
Nandra et al.~(1999). The point designated ``Int'' is for the profile
integrated over the whole observation. Points designated ``P1''
through ``P8'' are for shorter time intervals of the observation.
The panels differ by the model used
for the disk--black hole system just as in Figure 2.
\label{fig3}}

\figcaption[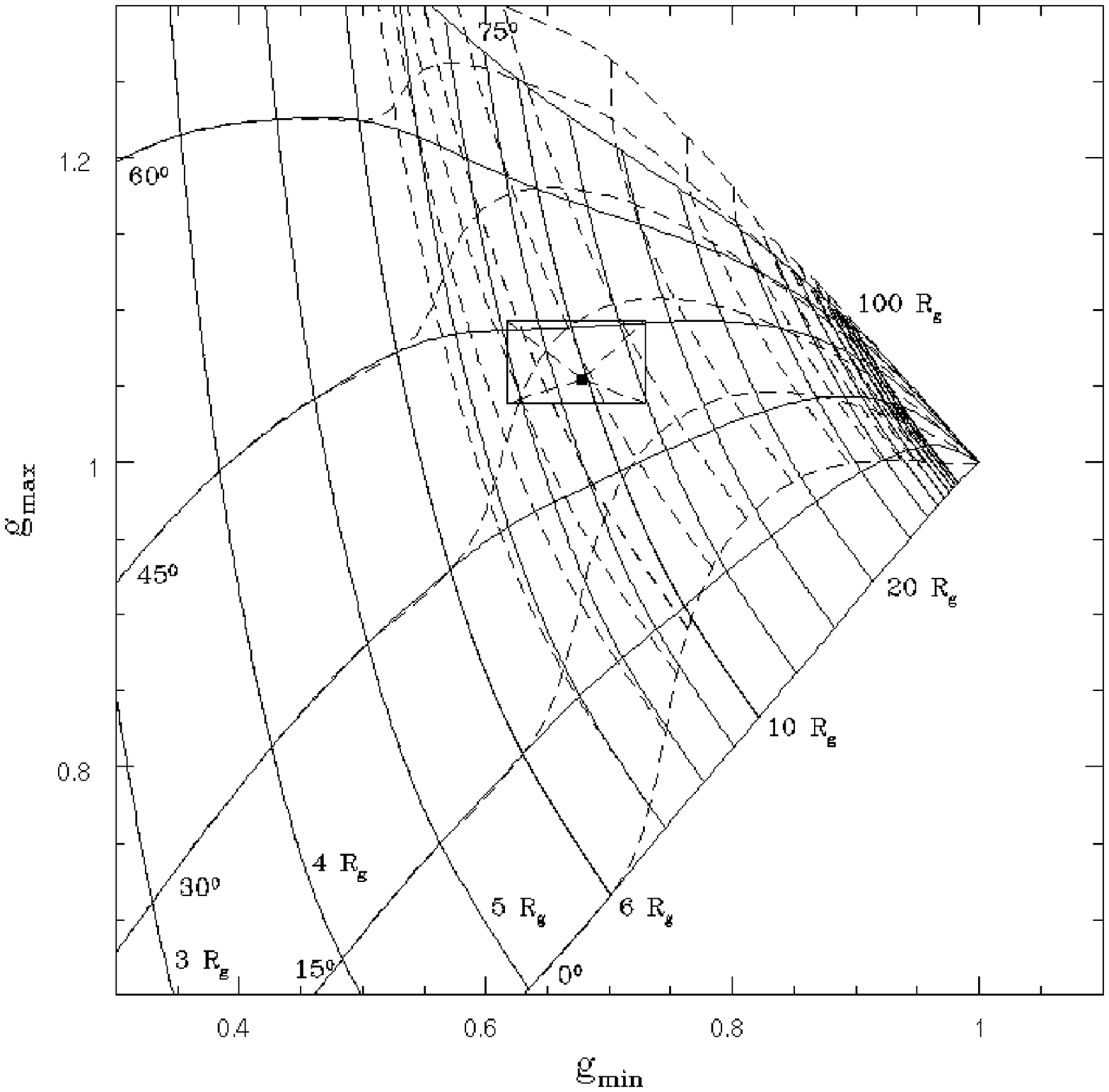, 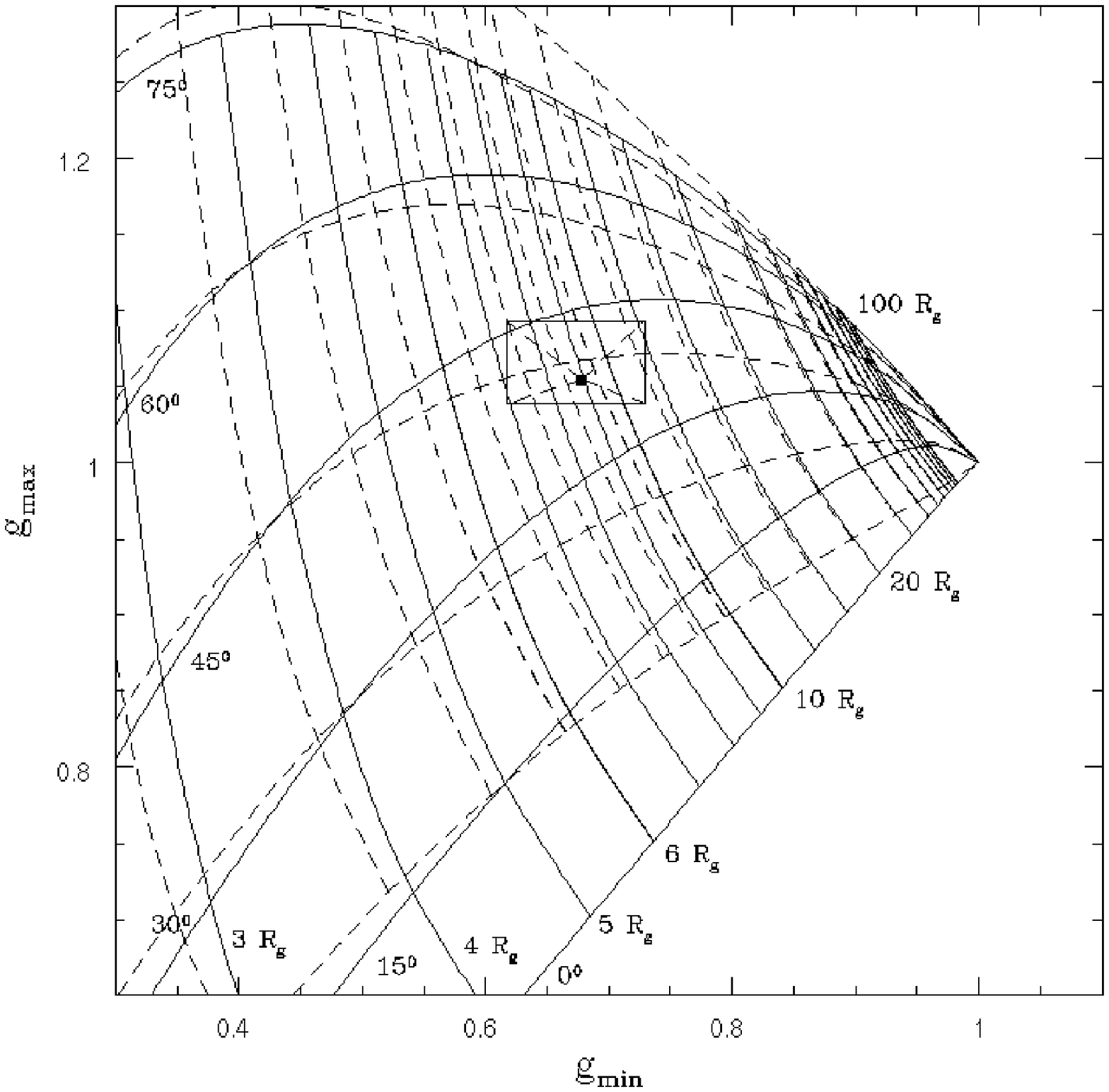]
{The results of continuum + line edge model fitting to the 
NGC~4151 data by Wang et al. (1999). All explanations and notations 
are the same as for Fig.~1 but 
using the non-linear ``sharp edge'' fitting model.
\label{fig5}}

\figcaption[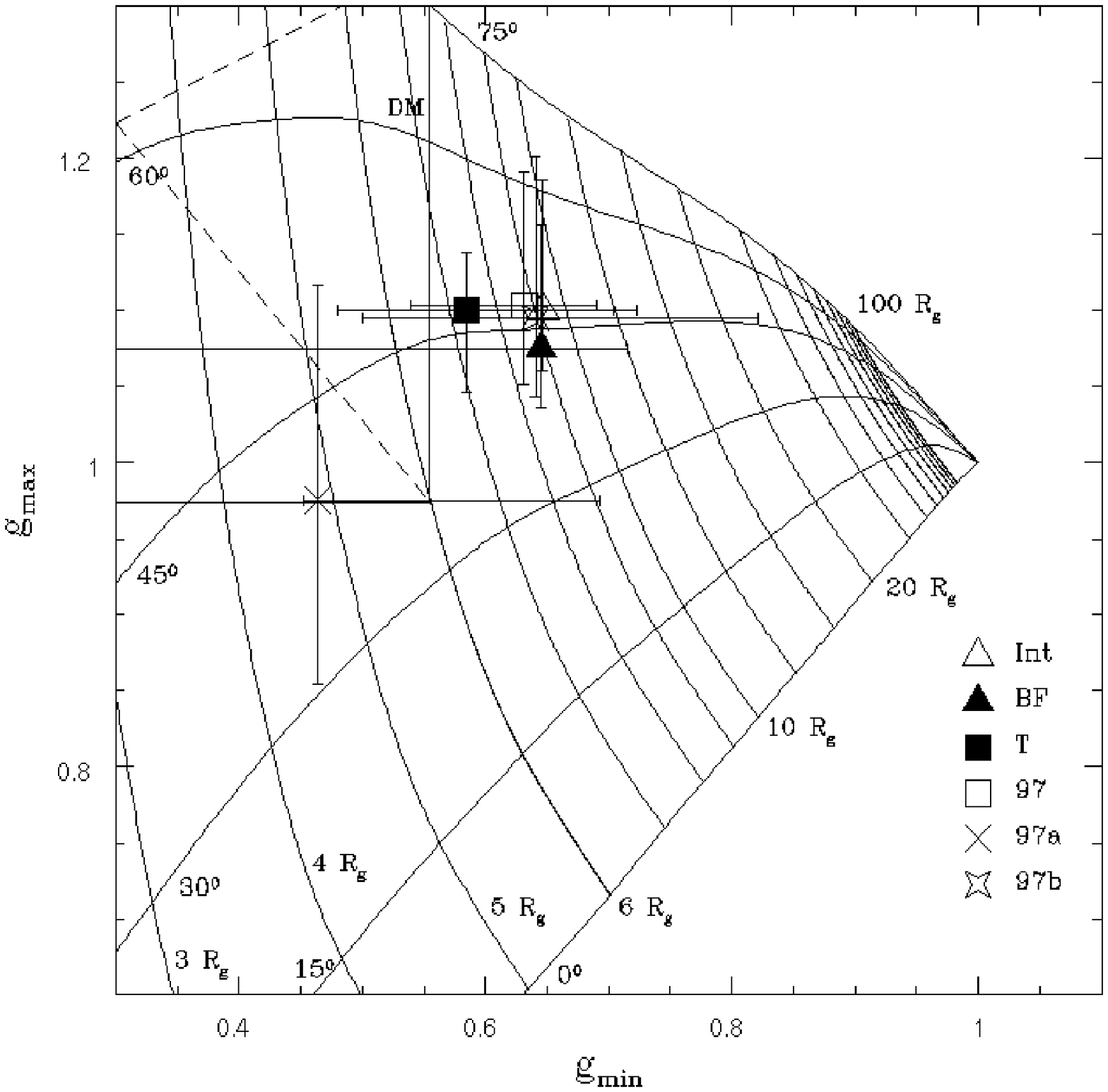, 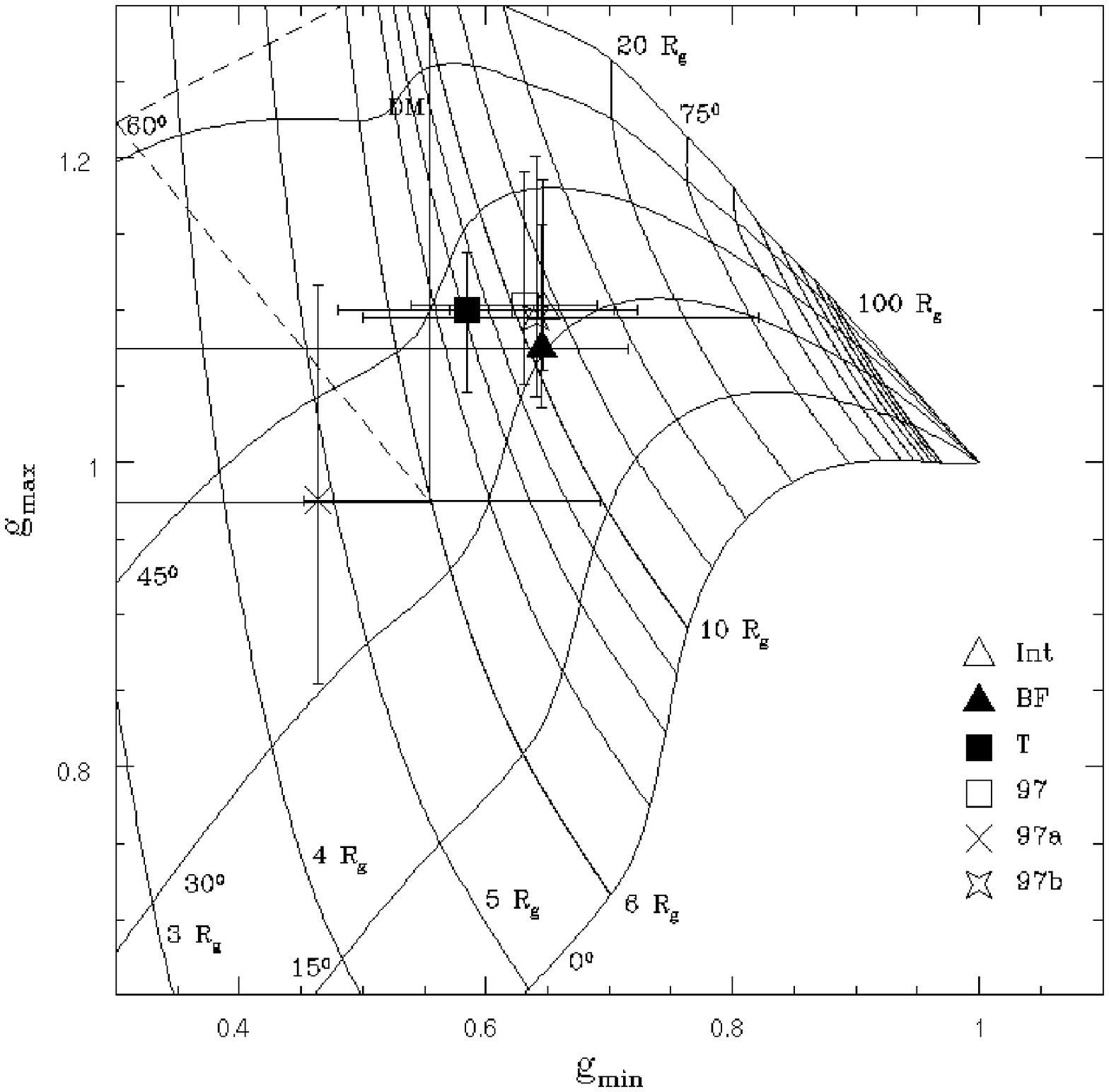, 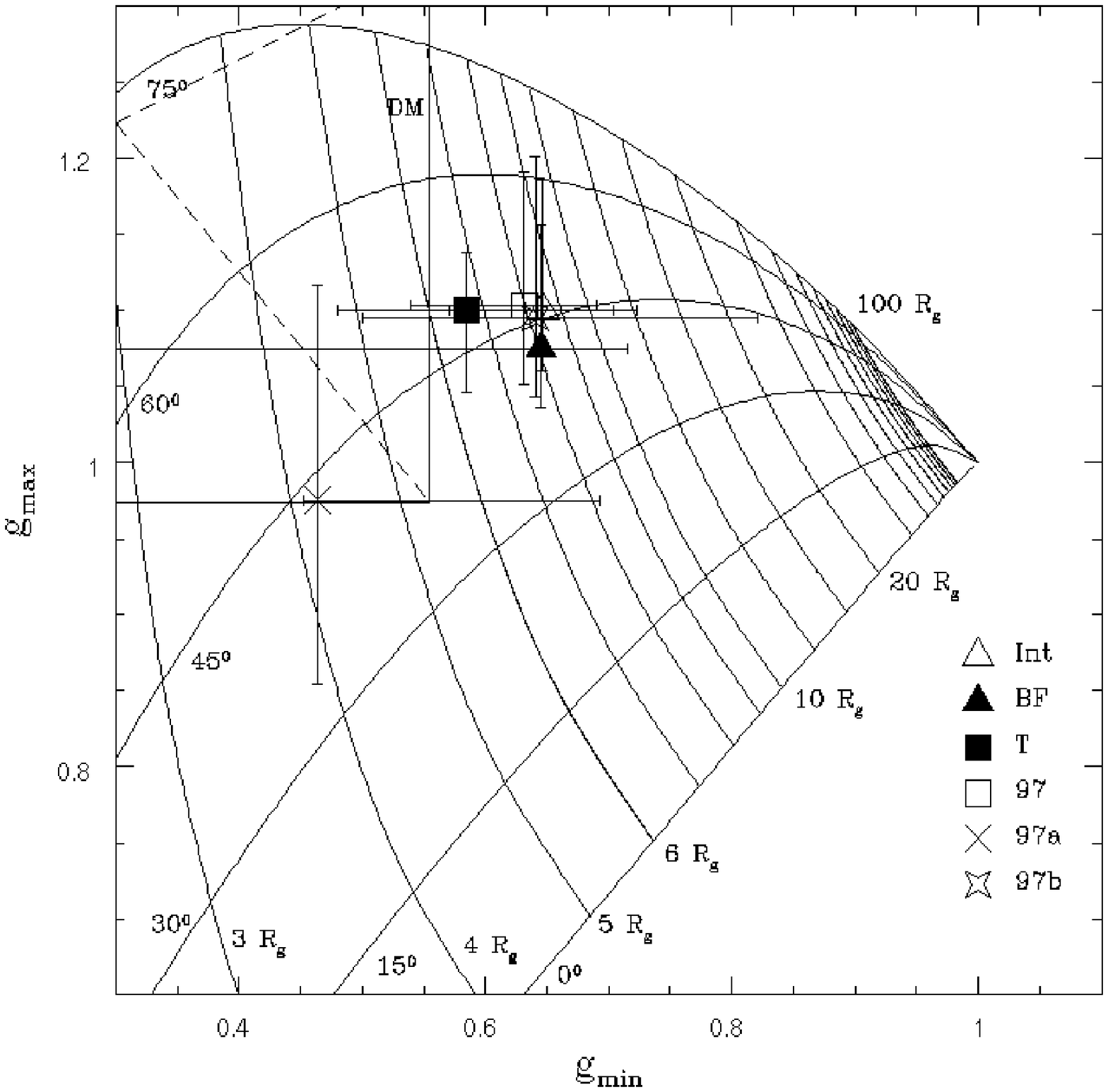, 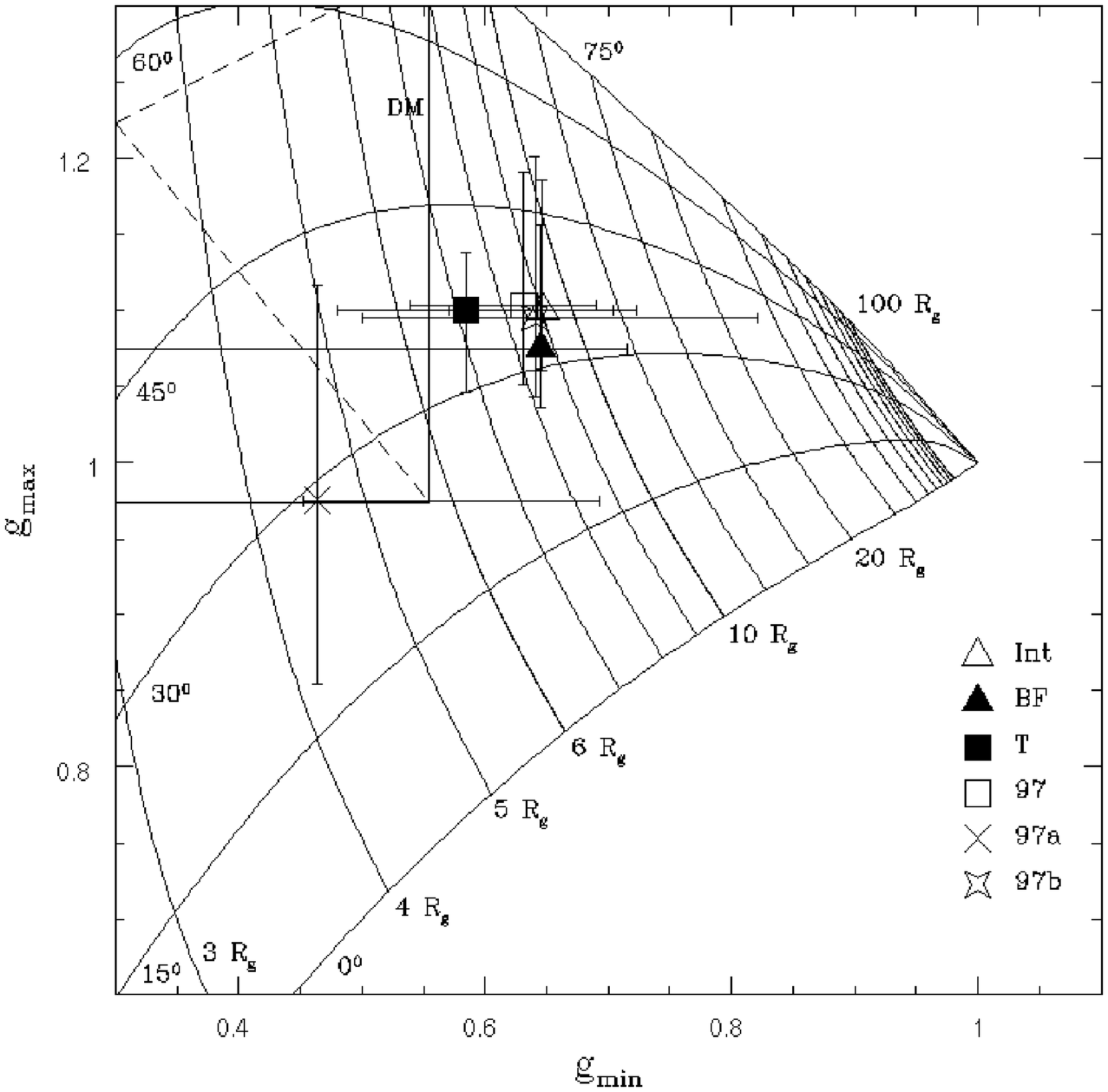]
{The results of continuum + line edge model fitting to the 
MCG-6-30-15 data. All explanations and notations are the same as 
for Fig.~2 but using the non-linear ``sharp edge'' fitting model.
Only blue limits for the red edge in the case of DM and BF data are
obtained. \label{fig6}}

\figcaption[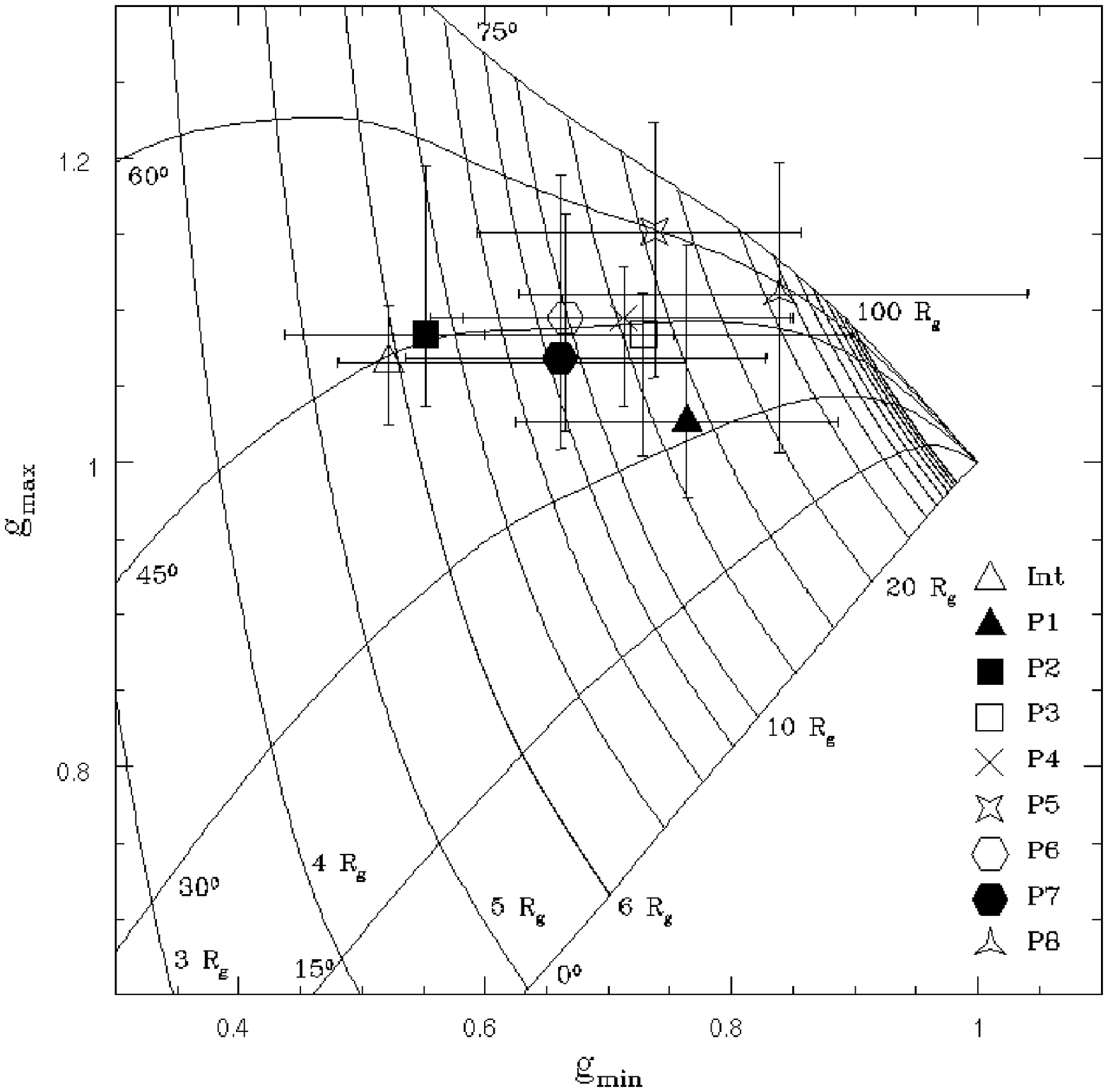, 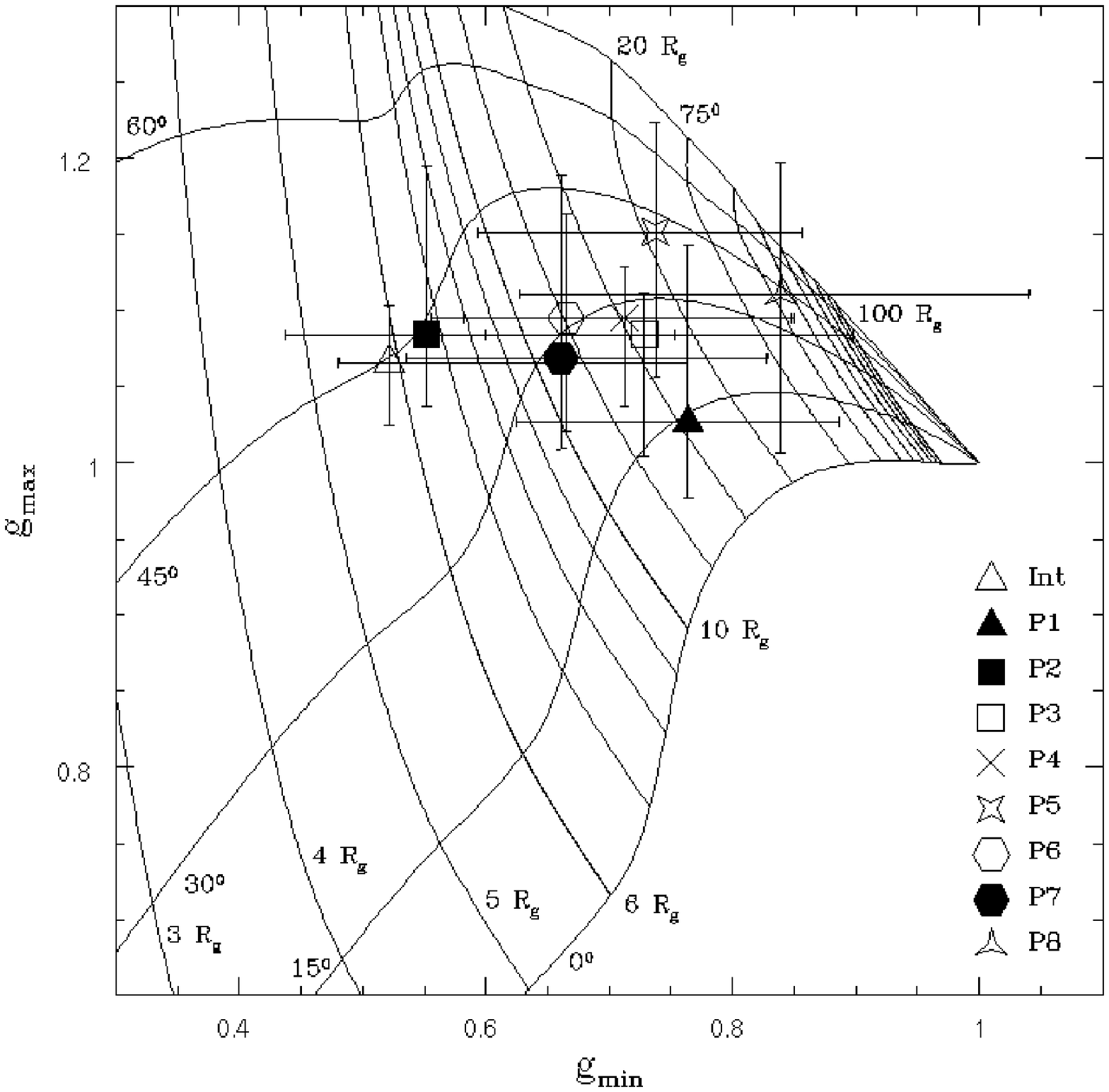, 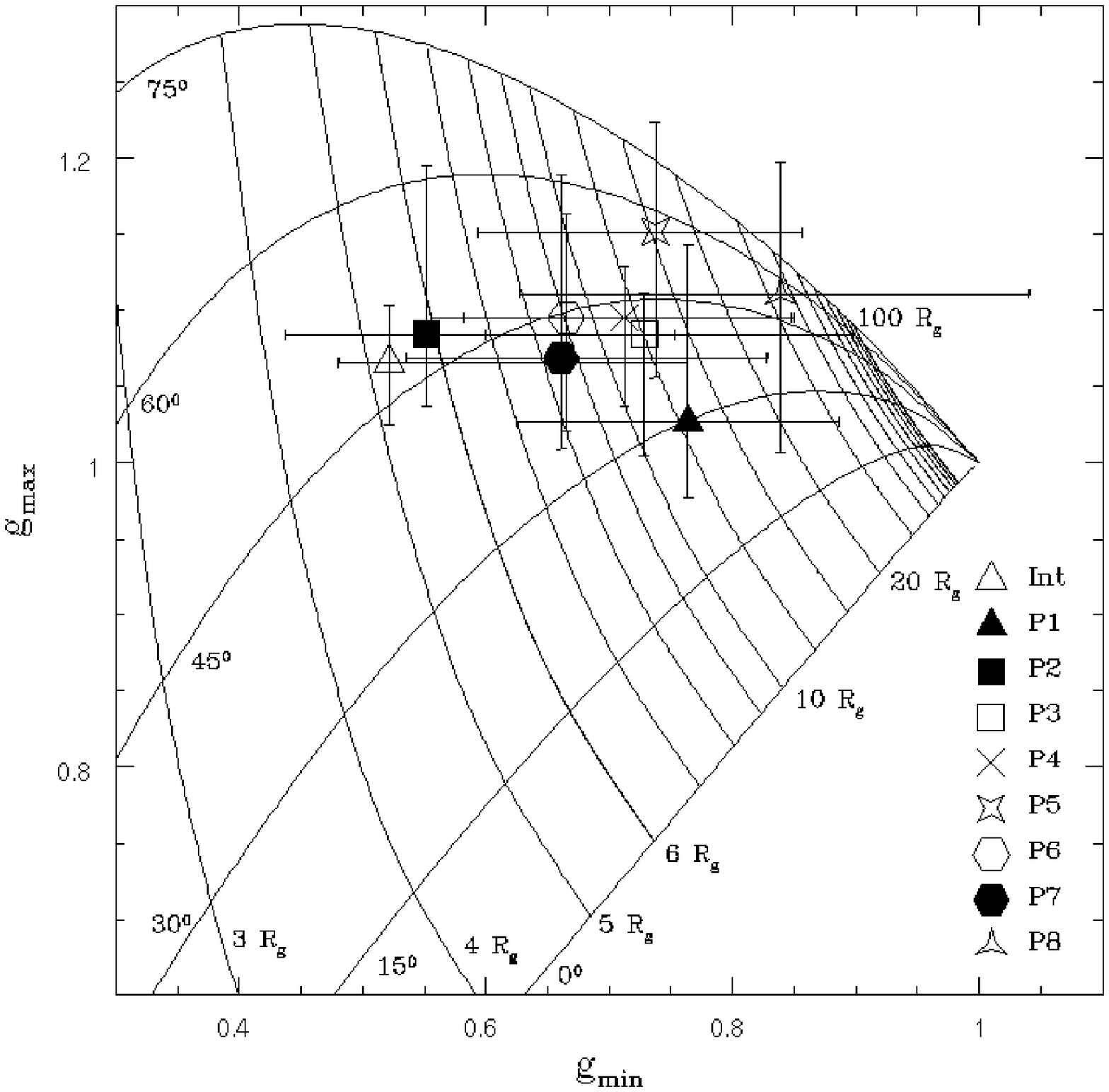, 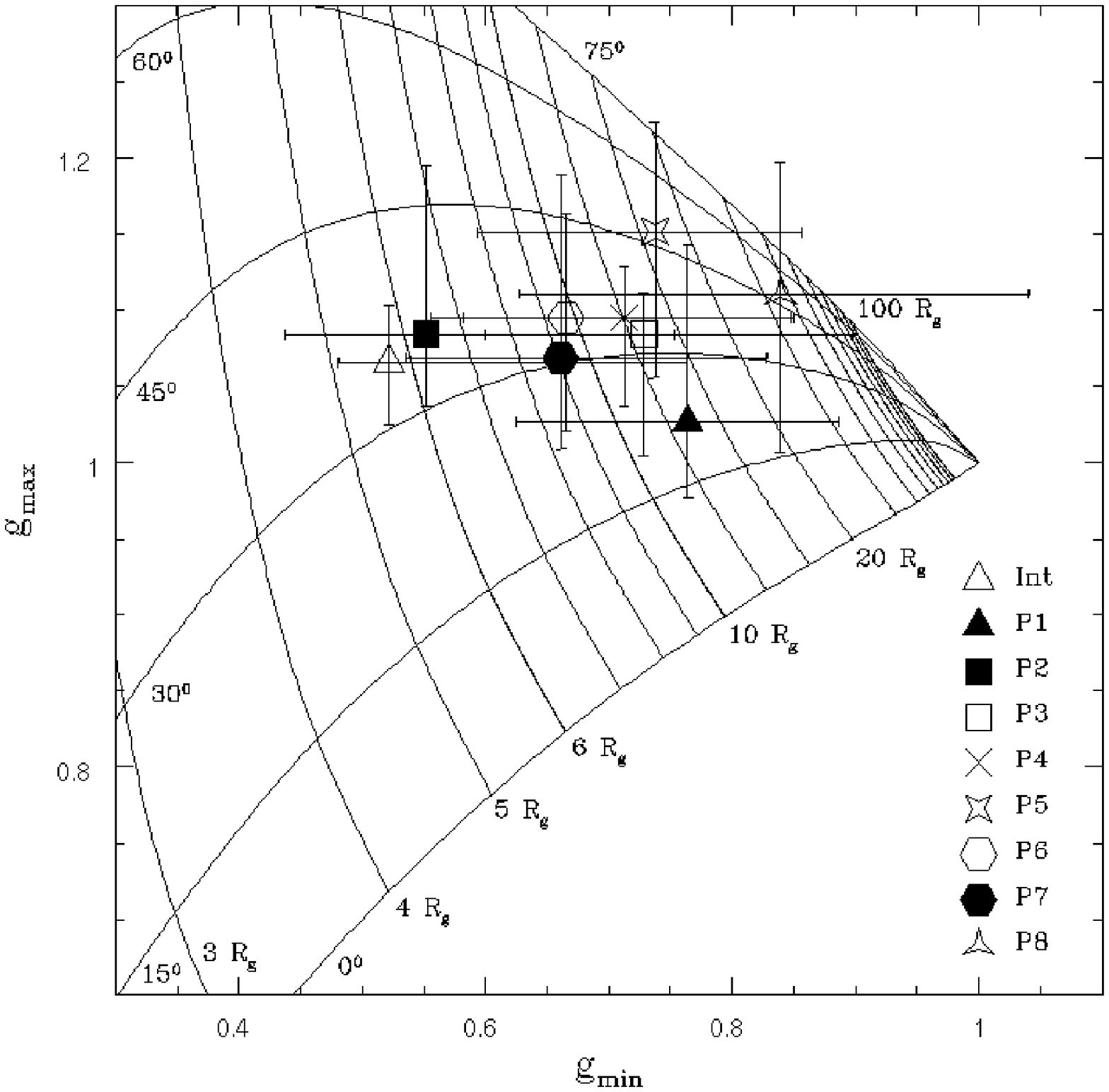]
{The results of continuum + line edge model fitting to the 
NGC~3516 data. All explanations and notations are the same as 
for Fig.~3 but using the non-linear ``sharp edge'' fitting model.
\label{fig7}}

\clearpage

\begin{figure}
\plotone{f1a.ps}
\end{figure}

\clearpage

\begin{figure}
\plotone{f1b.ps}
\end{figure}

\clearpage

\begin{figure}
\plotone{f2a.ps}
\end{figure}

\clearpage

\begin{figure}
\plotone{f2b.ps}
\end{figure}

\clearpage

\begin{figure}
\plotone{f2c.ps}
\end{figure}

\clearpage

\begin{figure}
\plotone{f2d.ps}
\end{figure}

\clearpage

\begin{figure}
\plotone{f3a.ps}
\end{figure}

\clearpage

\begin{figure}
\plotone{f3b.ps}
\end{figure}

\clearpage

\begin{figure}
\plotone{f3c.ps}
\end{figure}

\clearpage

\begin{figure}
\plotone{f3d.ps}
\end{figure}

\clearpage

\begin{figure}
\plotone{f4a.ps}
\end{figure}

\clearpage

\begin{figure}
\plotone{f4b.ps}
\end{figure}

\clearpage

\begin{figure}
\plotone{f5a.ps}
\end{figure}

\clearpage

\begin{figure}
\plotone{f5b.ps}
\end{figure}

\clearpage

\begin{figure}
\plotone{f5c.ps}
\end{figure}

\clearpage

\begin{figure}
\plotone{f5d.ps}
\end{figure}

\clearpage

\begin{figure}
\plotone{f6a.ps}
\end{figure}

\clearpage

\begin{figure}
\plotone{f6b.ps}
\end{figure}

\clearpage

\begin{figure}
\plotone{f6c.ps}
\end{figure}

\clearpage

\begin{figure}
\plotone{f6d.ps}
\end{figure}

\newpage

\begin{deluxetable}{rrrrrrr}  
\tablecolumns{7} 
\tablewidth{0pc}  
\tablecaption{Results of Fitting the Line Edges with Polynomial Model}
\tablehead{  
\colhead{Observation} & \colhead{Edge} & \colhead{Points} &  
\colhead{Approximation} & \colhead{keV} & \colhead{$g\pm 2\sigma$} &
\colhead{$\chi^2$/dof} \\
}
\startdata
Tanaka95 & Red & 1-16 & quadratic, MC around actual data & 3.67$\pm$0.25 & 0.57$\pm$0.08 & \nodata \\
Tanaka95 & Blue & 26-31 & quadratic, MC around actual data & 6.90$\pm$0.10 & 1.078$\pm$0.031 & \nodata \\
I96, Int & Red  & 1-29 & cubic, MC around actual data & 4.17$\pm$0.19 & 0.652$\pm$0.061 & 52.81/25 \\ 
I96, Int & Blue & 42-54 & quadratic, MC around actual data & 7.08$\pm$0.12 & 1.107$\pm$0.039 & 15.29/10 \\
I96, BF  & Red  & 1-32  & quadratic, MC around actual data  & 4.72$\pm$0.28 &  0.737$\pm$0.086 & 66.58/29 \\ 
I96, BF  & Blue & 35-45 & cubic, MC around actual data & 7.025$\pm$0.199 & 1.098$\pm$0.062 & 13.53/7 \\
I96, DM & Red &  1-22 & cubic, MC around actual data & 2.79$\pm$0.40 & 0.436$\pm$0.124 & 41.21/18 \\ 
I96, DM & Blue & 28-36 & quadratic, MC around actual data & 7.45$\pm$0.46 & 1.164$\pm$0.143 & 10.69/6 \\ 
I99, Int & Red & 1-6 & quadratic, MC around actual data & 3.79$\pm$0.30 & 0.59$\pm$0.09 & \nodata \\
I99, Int & Blue & 20-24 & quadratic, MC around actual data & 7.12$\pm$0.17 & 1.11$\pm$0.06 & \nodata \\ 
I99, a & Red  & 1-33 & quadratic, MC around actual data & 3.07$\pm$0.29 & 0.48$\pm$0.09 & \nodata \\
I99, a & Blue & 37-43 & quadratic, MC around actual data  & 6.24$\pm$0.23 & 0.975$\pm$0.07 & \nodata \\ 
I99, b & Red  & 1-34 & quadratic, MC around actual data & 4.29$\pm$0.47 & 0.67$\pm$0.15 & \nodata \\  
I99, b & Blue & 37-41 & quadratic, MC around actual data & 7.09$\pm$0.26 & 1.108$\pm$0.08 & \nodata \\ 
NGC 4151 & Red & 6-15 & quadratic, MC around actual data & 4.28$\pm$0.15 & 0.669$\pm$0.047 & \nodata \\  
NGC 4151 & Blue & 23-40 & quadratic, MC around actual data & 7.78$\pm$0.10 & 1.216$\pm$0.031 & \nodata \\
NGC 3516, Int & Red  & 1-19 & quadratic, MC around actual data & 3.53$\pm$0.19 & 0.55$\pm$0.06 & \nodata \\
NGC 3516, Int & Blue & 36-39 & quadratic, MC around actual data & 6.58$\pm$0.07 & 1.028$\pm$0.022 & \nodata \\
NGC 3516, P1 & Red & 18-36 & quadratic, MC around actual data & 4.58$\pm$0.32 & 0.72$\pm$0.10 & \nodata \\ 
NGC 3516, P1 & Blue & 37-43 & cubic, MC around actual data  & 6.51$\pm$0.07 & 1.02$\pm$0.02 & \nodata \\
NGC 3516, P2 & Red & 1-36 & quadratic, MC around actual data & 3.62$\pm$0.22 & 0.566$\pm$0.069 & \nodata \\ 
NGC 3516, P2 & Blue & 38-48 & cubic, MC around actual data & 7.14$\pm$0.23 & 1.12$\pm$0.07 & \nodata \\ 
NGC 3516, P3 & Red & 23-36 & cubic, MC around actual data & 4.70$\pm$0.15 & 0.734$\pm$0.047 & \nodata \\  
NGC 3516, P3 & Blue & 38-41 & quadratic, MC around actual data & 6.57$\pm$0.095 & 1.027$\pm$0.030 & \nodata \\ 
NGC 3516, P4 & Red & 1-31 & quadratic, MC around actual data & 3.94$\pm$0.52 & 0.616$\pm$0.081 & \nodata \\
NGC 3516, P4 & Blue & 38-47 & quadratic, MC around actual data & 6.99$\pm$0.19 & 1.092$\pm$0.059 & \nodata \\ 
NGC 3516, P5 & Red & 1-30 & cubic, MC around actual data & 4.73$\pm$0.20 & 0.739$\pm$0.063 & \nodata \\ 
NGC 3516, P5 & Blue & 38-48 & cubic, MC around actual data & 6.89$\pm$0.20 & 1.077$\pm$0.063 & \nodata \\
NGC 3516, P6 & Red & 19-30 & quadratic, MC around actual data & 4.38$\pm$0.24 & 0.6844$\pm$0.075 & \nodata \\ 
NGC 3516, P6 & Blue & 39-48 & quadratic, MC around actual data & 7.24$\pm$0.33 & 1.131$\pm$0.103 & \nodata \\
NGC 3516, P7 & Red & 18-36 & cubic, MC around actual data & 4.18$\pm$0.15 & 0.653$\pm$0.047 & \nodata \\ 
NGC 3516, P7 & Blue & 38-46 & quadratic, MC around actual data & 6.97$\pm$0.15 & 1.089$\pm$0.047 & \nodata \\
NGC 3516, P8 & Red & 1-36 & cubic, MC around actual data & 3.44$\pm$0.25 & 0.538$\pm$0.078 & \nodata \\ 
NGC 3516, P8 & Blue & 40-48 & quadratic, MC around actual data & 7.17$\pm$0.36 & 1.120$\pm$0.112 & \nodata \\
\enddata
\tablecomments{MC is short for Monte-Carlo; I96 in the first column refers to the data from Iwasawa
et al.~(1996) for Intermediate (Int), Bright Flare (BF),and Deep Minimum (DM) data sets; 
I99 refers to the data from Iwasawa et al.~(1999) for the whole observation (Int), for time 
intervals (a) and (b); Tanaka95  to the data from Tanaka et al. (1999); NGC 4151 data are 
from Wang et al.~(1999); NGC 3516 data are from Nandra et al. (1999) for the mean line profile 
(Int) and for 8 time intervals (P1-P8); the last column contains $\chi^2$ averaged over Monte-Carlo
realizations.}
\end{deluxetable}

\newpage

\begin{deluxetable}{rrrrrrr}  
\tablecolumns{7} 
\tablewidth{0pc}  
\tablecaption{Results of Fitting the Line Edges with Nonlinear Model}
\tablehead{  
\colhead{Observation} & \colhead{Edge} & \colhead{Points} &  
\colhead{Approximation} & \colhead{keV} & \colhead{$g\pm 2\sigma$} &  
\colhead{$\chi^2$/dof} \\
} 
\startdata
Tanaka95 & Red  & 1-16 & linear, MC around actual data  & 3.79$\pm$0.36 & 0.592$\pm$0.112 & 35.26/14 \\ 
Tanaka95 & Red  & 1-16 & linear, actual best fit     &  3.74 & \nodata & 16.95/14 \\
Tanaka95 & Blue & 26-34 & quadratic, MC around actual data & 6.99$\pm$0.15 & 1.092$\pm$0.046 & 15.96/6 \\ %plot
Tanaka95 & Blue & 26-34 & quadratic, actual best fit & 7.04 & \nodata & 8.37/6 \\
I96, Int & Red  & 1-29 & linear, MC around actual data  &   4.14$\pm$0.24 & 0.647$\pm$0.076 \\   % plot
I96, Int & Red  & 1-29 & linear, actual best fit       &  4.14 & \nodata  & 25.50/27 \\
I96, Int & Blue & 42-56 & quadratic, MC around actual data & 7.20$\pm$0.21 & 1.123$\pm$0.063 \\ % plot 
I96, Int & Blue & 42-56 & quadratic, actual best fit & 7.04 & \nodata & 5.01/12 \\
I96, BF  & Red  & 1-33  & linear, MC around actual data & 4.88$\pm$0.93 & 0.763$\pm$0.290 \\  % plot
I96, BF  & Red  & 1-33  & linear, actual best fit & 4.13 & \nodata & 43.71/31 \\
I96, BF  & Red  & \nodata  & n-points $\chi^2$ estimate\tablenotemark{a} & 4.58 & \nodata & \nodata \\
I96, BF  & Blue & 34-45 & quadratic, MC around actual data & 7.02$\pm$0.19 & 1.096$\pm$0.060 \\ % plot
I96, BF  & Blue & 34-45 & quadratic, actual best fit & 6.88  & \nodata & 6.05/9 \\
I96, DM & Blue & 28-36 & linear, MC around actual data & 7.37$\pm$0.57 & 1.152$\pm$0.178 &\nodata \\ % plot
I96, DM & Blue & 28-36 & linear, actual best fit & 7.83 & \nodata & 3.75/7 \\
I96, DM & Red & \nodata & n-points $\chi^2$ estimate\tablenotemark{a} & 3.55 & \nodata & \nodata \\
I99, Int & Red  & 1-7 & quadratic, MC around actual data & 3.94$\pm$0.24 & 0.615$\pm$0.076 & 5.91/4 \\ % plot
I99, Int & Red  & 1-7 & quadratic, actual best fit  & 4.04 & \nodata & 0.83/4 \\
I99, Int & Blue & 20-25 & quadratic, MC around actual data & 7.17$\pm$0.22 & 1.121$\pm$0.070 & 5.16/3 \\ % plot
I99, Int & Blue & 20-25 & quadratic, actual best fit & 7.06 & \nodata & 1.65/3 \\
I99, a & Red  & 1-34 & linear, MC around actual data & 3.67$\pm$0.38 & 0.573$\pm$0.120 & 57.44/32 \\ % plot
I99, a & Red  & 1-34 & linear, actual best fit    & 2.97 & \nodata & 23.45/32 \\
I99, a & Blue & 37-43 & linear, MC around actual data  & 6.31$\pm$0.42 & 0.986$\pm$0.131 & 9.33/5 \\ % plot
I99, a & Blue & 37-43 & linear, actual best fit   & 6.24 & \nodata & 4.76/5 \\
I99, b & Red  & 1-34 & linear, MC around actual data & 4.23$\pm$0.52 & 0.661$\pm$0.161 & 60.48/32 \\  % plot
I99, b & Red  & 1-34 & linear, actual best fit    & 4.10 & \nodata & 25.74/32 \\
I99, b & Blue & 37-41 & linear, MC around actual data  & 7.18$\pm$0.25 & 1.122$\pm$0.079 & 3.84/3 \\ % plot
I99, b & Blue & 37-41 & linear, actual best fit   & 7.01 & \nodata & 0.800/3 \\
NGC 4151 & Red & 1-15 & linear, MC around actual data & 4.31$\pm$0.18 & 0.673$\pm$0.056 & 56.77/13 \\
NGC 4151 & Red & 1-15 & linear, actual best fit       & 4.34 & \nodata & 17.35/13 \\   % plot 
NGC 4151 & Blue & 23-40 & quadratic, MC around actual data & 6.82$\pm$0.087 & 1.066$\pm$0.027 & 67.65/15 \\ % plot
NGC 4151 & Blue & 23-40 & quadratic, actual best fit    & 6.75 & \nodata & 20.73/15 \\
NGC 3516, Int & Red  & 5-24 & linear, MC around actual data & 3.98$\pm$0.45 & 0.622$\pm$0.141 & 81.33/18 \\ % plot
NGC 3516, Int & Red  & 5-24 & linear, actual best fit    & 3.34 & \nodata & 27.54/18 \\
NGC 3516, Int & Blue & 34-45 & linear, MC around actual data  & 6.81$\pm$0.13 & 1.064$\pm$0.039 & 41.72/10 \\ % plot
NGC 3516, Int & Blue & 34-45 & linear, actual best fit   & 6.82 & \nodata & 16.26/10 \\
NGC 3516, P1 & Red & 18-36 & linear, MC around actual data & 4.83$\pm$0.42 & 0.756$\pm$0.131 & 51.67/17 \\  % plot
NGC 3516, P1 & Red & 18-36 & linear, actual best fit    & 4.89 & \nodata & 21.51/17 \\
NGC 3516, P1 & Blue & 38-48 & linear, MC around actual data  & 6.78$\pm$0.27 & 1.060$\pm$0.083 & 16.68/9 \\ % plot
NGC 3516, P1 & Blue & 38-48 & linear, actual best fit     & 6.57 & \nodata & 3.85/9 \\
\tablebreak
NGC 3516, P2 & Red & 1-36 & linear, MC around actual data & 3.81$\pm$0.50 & 0.595$\pm$0.158 & 96.05/34 \\  % plot
NGC 3516, P2 & Red & 1-36 & linear, actual best fit    & 3.53 & \nodata & 30.98/34 \\
NGC 3516, P2 & Blue & 37-48 & linear, MC around actual data  & 7.14$\pm$0.25 & 1.116$\pm$0.079 & 21.62/10 \\ % plot
NGC 3516, P2 & Blue & 37-48 & linear, actual best fit     & 6.94 & \nodata & 6.13/10 \\
NGC 3516, P3 & Red & 23-36 & linear, MC around actual data & 4.79$\pm$0.48 & 0.749$\pm$0.149 & 27.89/12 \\  % plot 
NGC 3516, P3 & Red & 23-36 & linear, actual best fit    & 4.66 & \nodata & 4.86/12 \\
NGC 3516, P3 & Blue & 37-48 & quadratic, MC around actual data & 6.77$\pm$0.17 & 1.058$\pm$0.054 & 19.78/9 \\ % plot
NGC 3516, P3 & Blue & 37-48 & quadratic, actual best fit & 6.94 & \nodata & 5.07/9 \\
NGC 3516, P4 & Red & 13-31 & linear, MC around actual data & 4.50$\pm$0.47 & 0.703$\pm$0.147 & 41.52/17 \\ % plot
NGC 3516, P4 & Red & 13-31 & linear, actual best fit          & 4.56  & \nodata & 10.67/17 \\
NGC 3516, P4 & Blue & 37-48 & quadratic, MC around actual data & 6.93$\pm$0.147 & 1.083$\pm$0.046 & 23.96/9 \\  % plot
NGC 3516, P4 & Blue & 37-48 & quadratic, actual best fit       & 7.01 & \nodata & 10.74/9 \\
NGC 3516, P5 & Red & 1-30 & quadratic, MC around actual data & 4.64$\pm$0.42 & 0.725$\pm$0.131 & 76.6/27  \\  % plot
NGC 3516, P5 & Red & 1-30 & quadratic, actual best fit       & 4.72 & \nodata & 25.65/27  \\
NGC 3516, P5 & Blue & 38-48 & linear, MC around actual data & 7.29$\pm$0.27 & 1.140$\pm$0.084 & 21.48/9 \\  % plot
NGC 3516, P5 & Blue & 38-48 & linear, actual best fit & 7.37 & \nodata & 8.42/9 \\
NGC 3516, P6 & Red & 1-30 & linear, MC around actual data & 4.58$\pm$0.42 & 0.715$\pm$0.133 & 77.86/28 \\   % plot
NGC 3516, P6 & Red & 1-30 & linear, actual best fit & 4.26 & \nodata & 27.93/28 \\
NGC 3516, P6 & Blue & 38-48 & quadratic, MC around actual data & 6.99$\pm$0.23 & 1.092$\pm$0.071 & 15.68/8 \\ % plot
NGC 3516, P6 & Blue & 38-48 & quadratic, actual best fit & 7.01 & \nodata & 4.26/8 \\
NGC 3516, P7 & Red & 1-36 & linear, MC around actual data & 4.36$\pm$0.47 & 0.682$\pm$0.146 & 98.34/34   \\ % plot
NGC 3516, P7 & Red & 1-36 & linear, actual best fit & 4.23 & \nodata & 36.99/34    \\
NGC 3516, P7 & Blue & 38-48 & quadratic, MC around actual data & 7.03$\pm$0.29 & 1.099$\pm$0.090 & 17.32/8  \\ % plot
NGC 3516, P7 & Blue & 38-48 & quadratic, actual best fit & 6.84 & \nodata & 6.16/8  \\
NGC 3516, P8 & Red & 1-37 & quadratic, MC around actual data & 5.33$\pm$0.66 & 0.834$\pm$0.207 & 100.5/34 \\ % plot
NGC 3516, P8 & Red & 1-37 & quadratic, actual best fit    & 5.37 & \nodata & 40.96/34 \\
NGC 3516, P8 & Blue & 39-47 & quadratic, MC around actual data & 7.05$\pm$0.30 & 1.102$\pm$0.095 & 15.00/6 \\ 
NGC 3516, P8 & Blue & 39-47 & quadratic, actual best fit       & 7.11 & \nodata & 6.77/6 \\ % plot
\enddata
\tablenotetext{a}{Blue conservative limit on the red edge from the n-points $\chi^2$ rejection
method}
\tablecomments{MC is short for Monte-Carlo; I96 in the first column refers to the data from Iwasawa
et al.~(1996) for Intermediate (Int), Bright Flare (BF),and Deep Minimum (DM) data sets; 
I99 refers to the data from Iwasawa et al.~(1999) for the whole observation (Int), for time 
intervals (a) and (b); Tanaka95 refers to the data from Tanaka et al. (1999); NGC 4151 data are 
from Wang et al.~(1999); NGC 3516 data are from Nandra et al. (1999) for the mean line profile 
(Int) and for 8 time intervals (P1-P8); the last column contains $\chi^2$ averaged over Monte-Carlo
realizations.}
\end{deluxetable}

\end{document}